 \input pictex.tex   

\immediate\write10{Package DCpic 2002/05/16 v4.0}

\catcode`!=11 

\newcount\aux%
\newcount\auxa%
\newcount\auxb%
\newcount\m%
\newcount\n%
\newcount\x%
\newcount\y%
\newcount\xl%
\newcount\yl%
\newcount\d%
\newcount\dnm%
\newcount\xa%
\newcount\xb%
\newcount\xmed%
\newcount\xc%
\newcount\xd%
\newcount\ya%
\newcount\yb%
\newcount\ymed%
\newcount\yc%
\newcount\yd
\newcount\expansao%
\newcount\tipografo
\newcount\distanciaobjmor
\newcount\tipoarco
\newif\ifpara%
\newbox\caixa%
\newbox\caixaaux%
\newif\ifnvazia%
\newif\ifvazia%
\newif\ifcompara%
\newif\ifdiferentes%
\newcount\xaux%
\newcount\yaux%
\newcount\guardaauxa%
\newcount\alt%
\newcount\larg%
\newcount\prof%
\newcount\auxqx
\newcount\auxqy
\newif\ifajusta%
\newif\ifajustadist
\def\objPartida{}%
\def\objChegada{}%
\def\objNulo{}%


\def\!vazia{:}

\def\!pilhanvazia#1{\let\arg=#1%
\if:\arg\ \nvaziafalse\vaziatrue \else \nvaziatrue\vaziafalse\fi}

\def\!coloca#1#2{\edef\pilha{#1.#2}}

\def\!guarda(#1)(#2,#3)(#4,#5,#6){\def\id{#1}%
\xaux=#2%
\yaux=#3%
\alt=#4%
\larg=#5%
\prof=#6%
}

\def\!topaux#1.#2:{\!guarda#1}
\def\!topo#1{\expandafter\!topaux#1}

\def\!popaux#1.#2:{\def\pilha{#2:}}
\def\!retira#1{\expandafter\!popaux#1}

\def\!comparaaux#1#2{\let\argA=#1\let\argB=#2%
\ifx\argA\argB\comparatrue\diferentesfalse\else\comparafalse\diferentestrue\fi}

\def\!compara#1#2{\!comparaaux{#1}{#2}}

\def\!absoluto#1#2{\n=#1%
  \ifnum \n > 0
    #2=\n
  \else
    \multiply \n by -1
    #2=\n
  \fi}






\def\!ajusta#1#2#3#4#5#6{\aux=#5%
  \let\auxobj=#6%
  \ifcase \tipografo    
    \ifnum\number\aux=10 
      \ajustadisttrue 
    \else
      \ajustadistfalse  
    \fi
  \else  
   \ajustadistfalse
  \fi
  \ifajustadist
   \let\pilhaaux=\pilha%
   \loop%
     \!topo{\pilha}%
     \!retira{\pilha}%
     \!compara{\id}{\auxobj}%
     \ifcompara\nvaziafalse \else\!pilhanvazia\pilha \fi%
     \ifnvazia%
   \repeat%
   \let\pilha=\pilhaaux%
   \ifvazia%
    \ifdiferentes%
     \larg=1310720
     \prof=655360%
     \alt=655360%
    \fi%
   \fi%
   \divide\larg by 131072
   \divide\prof by 65536
   \divide\alt by 65536
   \ifnum\number\y=\number\yl
    \advance\larg by 3
    \ifnum\number\larg>\aux
     #5=\larg
    \fi
   \else
    \ifnum\number\x=\number\xl
     \ifnum\number\yl>\number\y
      \ifnum\number\alt>\aux
       #5=\alt
      \fi
     \else
      \advance\prof by 5
      \ifnum\number\prof>\aux
       #5=\prof
      \fi
     \fi
    \else
     \auxqx=\x
     \advance\auxqx by -\xl
     \!absoluto{\auxqx}{\auxqx}%
     \auxqy=\y
     \advance\auxqy by -\yl
     \!absoluto{\auxqy}{\auxqy}%
     \ifnum\auxqx>\auxqy
      \ifnum\larg<10
       \larg=10
      \fi
      \advance\larg by 3
      #5=\larg
     \else
      \ifnum\yl>\y
       \ifnum\larg<10
        \larg=10
       \fi
      \advance\alt by 6
       #5=\alt
      \else
      \advance\prof by 11
       #5=\prof
      \fi
     \fi
    \fi
   \fi
\fi} 

\def\!raiz#1#2{\n=#1%
  \m=1%
  \loop
    \aux=\m%
    \advance \aux by 1%
    \multiply \aux by \aux%
    \ifnum \aux < \n%
      \advance \m by 1%
      \paratrue%
    \else\ifnum \aux=\n%
      \advance \m by 1%
      \paratrue%
       \else\parafalse%
       \fi
    \fi
  \ifpara%
  \repeat
#2=\m}

\def\!ucoord#1#2#3#4#5#6#7{\aux=#2%
  \advance \aux by -#1%
  \multiply \aux by #4%
  \divide \aux by #5%
  \ifnum #7 = -1 \multiply \aux by -1 \fi%
  \advance \aux by #3%
#6=\aux}

\def\!quadrado#1#2#3{\aux=#1%
  \advance \aux by -#2%
  \multiply \aux by \aux%
#3=\aux}

\def\!distnomemor#1#2#3#4#5#6{\setbox0=\hbox{#5}%
  \aux=#1
  \advance \aux by -#3
  \ifnum \aux=0
     \aux=\wd0 \divide \aux by 131072
     \advance \aux by 3
     #6=\aux
  \else
     \aux=#2
     \advance \aux by -#4
     \ifnum \aux=0
        \aux=\ht0 \advance \aux by \dp0 \divide \aux by 131072
        \advance \aux by 3
        #6=\aux%
     \else
     #6=3
     \fi
   \fi
}


\def\!begindc#1[#2]{\beginpicture 
  \let\pilha=\!vazia
  \setcoordinatesystem units <1pt,1pt>
  \expansao=#2
  \ifcase #1
    \distanciaobjmor=10
    \tipoarco=0         
    \tipografo=0        
  \or
    \distanciaobjmor=2
    \tipoarco=0         
    \tipografo=1        
  \or
    \distanciaobjmor=1
    \tipoarco=2         
    \tipografo=2        
  \or
    \distanciaobjmor=8
    \tipoarco=0         
    \tipografo=3        
  \or
    \distanciaobjmor=8
    \tipoarco=2         
    \tipografo=4        
  \fi}


\def\!morxy(#1,#2){%
  \!ifnextchar({\!morxyl{#1}{#2}}{\!morObjB{#1}{#2}}}
\def\!morxyl#1#2(#3,#4){%
  \!ifnextchar[{\!mora{#1}{#2}{#3}{#4}}{\!mora{#1}{#2}{#3}{#4}[\number\distanciaobjmor,\number\distanciaobjmor]}}%
\def\!morObjA#1{%
 \let\pilhaaux=\pilha%
 \def\objPartida{#1}%
 \loop%
    \!topo\pilha%
    \!retira\pilha%
    \!compara{\id}{\objPartida}%
    \ifcompara \nvaziafalse \else \!pilhanvazia\pilha \fi%
   \ifnvazia%
 \repeat%
 \ifvazia%
  \ifdiferentes%
   Error: Incorrect label specification%
   \xaux=1%
   \yaux=1%
  \fi%
 \fi%
 \let\pilha=\pilhaaux%
 \!ifnextchar({\!morxyl{\number\xaux}{\number\yaux}}{\!morObjB{\number\xaux}{\number\yaux}}}
\def\!morObjB#1#2#3{%
  \x=#1
  \y=#2
 \def\objChegada{#3}%
 \let\pilhaaux=\pilha%
 \loop
    \!topo\pilha %
    \!retira\pilha%
    \!compara{\id}{\objChegada}%
    \ifcompara \nvaziafalse \else \!pilhanvazia\pilha \fi
   \ifnvazia
 \repeat
 \ifvazia
  \ifdiferentes%
   Error: Incorrect label specification
   \xaux=\x%
   \advance\xaux by \x%
   \yaux=\y%
   \advance\yaux by \y%
  \fi
 \fi
 \let\pilha=\pilhaaux
 \!ifnextchar[{\!mora{\number\x}{\number\y}{\number\xaux}{\number\yaux}}{\!mora{\number\x}{\number\y}{\number\xaux}{\number\yaux}[\number\distanciaobjmor,\number\distanciaobjmor]}}
\def\!mora#1#2#3#4[#5,#6]#7{%
  \!ifnextchar[{\!morb{#1}{#2}{#3}{#4}{#5}{#6}{#7}}{\!morb{#1}{#2}{#3}{#4}{#5}{#6}{#7}[1,\number\tipoarco] }}
\def\!morb#1#2#3#4#5#6#7[#8,#9]{\x=#1%
  \y=#2%
  \xl=#3%
  \yl=#4%
  \multiply \x by \expansao%
  \multiply \y by \expansao%
  \multiply \xl by \expansao%
  \multiply \yl by \expansao%
  \!quadrado{\number\x}{\number\xl}{\auxa}%
  \!quadrado{\number\y}{\number\yl}{\auxb}%
  \d=\auxa%
  \advance \d by \auxb%
  \!raiz{\d}{\d}%
  \auxa=#5
  \!compara{\objNulo}{\objPartida}%
  \ifdiferentes
   \!ajusta{\x}{\xl}{\y}{\yl}{\auxa}{\objPartida}%
   \ajustatrue
   \def\objPartida{}
  \fi
  \guardaauxa=\auxa
  \!ucoord{\number\x}{\number\xl}{\number\x}{\auxa}{\number\d}{\xa}{1}%
  \!ucoord{\number\y}{\number\yl}{\number\y}{\auxa}{\number\d}{\ya}{1}%
  \auxa=\d%
  \auxb=#6
  \!compara{\objNulo}{\objChegada}%
  \ifdiferentes
   \!ajusta{\x}{\xl}{\y}{\yl}{\auxb}{\objChegada}%
   \def\objChegada{}
  \fi
  \advance \auxa by -\auxb%
  \!ucoord{\number\x}{\number\xl}{\number\x}{\number\auxa}{\number\d}{\xb}{1}%
  \!ucoord{\number\y}{\number\yl}{\number\y}{\number\auxa}{\number\d}{\yb}{1}%
  \xmed=\xa%
  \advance \xmed by \xb%
  \divide \xmed by 2
  \ymed=\ya%
  \advance \ymed by \yb%
  \divide \ymed by 2
  \!distnomemor{\number\x}{\number\y}{\number\xl}{\number\yl}{#7}{\dnm}%
  \!ucoord{\number\y}{\number\yl}{\number\xmed}{\number\dnm}{\number\d}{\xc}{-#8}%
  \!ucoord{\number\x}{\number\xl}{\number\ymed}{\number\dnm}{\number\d}{\yc}{#8}%
\ifcase #9  
  \arrow <4pt> [.2,1.1] from {\xa} {\ya} to {\xb} {\yb}
\or  
  \setdashes
  \arrow <4pt> [.2,1.1] from {\xa} {\ya} to {\xb} {\yb}
  \setsolid
\or  
  \setlinear
  \plot {\xa} {\ya}  {\xb} {\yb} /
\or  
  \auxa=\guardaauxa
  \advance \auxa by 3%
 \!ucoord{\number\x}{\number\xl}{\number\x}{\number\auxa}{\number\d}{\xa}{1}%
 \!ucoord{\number\y}{\number\yl}{\number\y}{\number\auxa}{\number\d}{\ya}{1}%
 \!ucoord{\number\y}{\number\yl}{\number\xa}{3}{\number\d}{\xd}{-1}%
 \!ucoord{\number\x}{\number\xl}{\number\ya}{3}{\number\d}{\yd}{1}%
  \arrow <4pt> [.2,1.1] from {\xa} {\ya} to {\xb} {\yb}
  \circulararc -180 degrees from {\xa} {\ya} center at {\xd} {\yd}
\or  
  \auxa=3
 \!ucoord{\number\y}{\number\yl}{\number\xa}{\number\auxa}{\number\d}{\xmed}{-1}%
 \!ucoord{\number\x}{\number\xl}{\number\ya}{\number\auxa}{\number\d}{\ymed}{1}%
 \!ucoord{\number\y}{\number\yl}{\number\xa}{\number\auxa}{\number\d}{\xd}{1}%
 \!ucoord{\number\x}{\number\xl}{\number\ya}{\number\auxa}{\number\d}{\yd}{-1}%
  \arrow <4pt> [.2,1.1] from {\xa} {\ya} to {\xb} {\yb}
  \setlinear
  \plot {\xmed} {\ymed}  {\xd} {\yd} /
\fi
\auxa=\xl
\advance \auxa by -\x%
\ifnum \auxa=0 
  \put {#7} at {\xc} {\yc}
\else
  \auxb=\yl
  \advance \auxb by -\y%
  \ifnum \auxb=0 \put {#7} at {\xc} {\yc}
  \else 
    \ifnum \auxa > 0 
      \ifnum \auxb > 0
        \ifnum #8=1
          \put {#7} [rb] at {\xc} {\yc}
        \else 
          \put {#7} [lt] at {\xc} {\yc}
        \fi
      \else
        \ifnum #8=1
          \put {#7} [lb] at {\xc} {\yc}
        \else 
          \put {#7} [rt] at {\xc} {\yc}
        \fi
      \fi
    \else
      \ifnum \auxb > 0 
        \ifnum #8=1
          \put {#7} [rt] at {\xc} {\yc}
        \else 
          \put {#7} [lb] at {\xc} {\yc}
        \fi
      \else
        \ifnum #8=1
          \put {#7} [lt] at {\xc} {\yc}
        \else 
          \put {#7} [rb] at {\xc} {\yc}
        \fi
      \fi
    \fi
  \fi
\fi
}

\def\modifplot(#1{\!modifqcurve #1}
\def\!modifqcurve(#1,#2){\x=#1%
  \y=#2%
  \multiply \x by \expansao%
  \multiply \y by \expansao%
  \!start (\x,\y)
  \!modifQjoin}
\def\!modifQjoin(#1,#2)(#3,#4){\x=#1%
  \y=#2%
  \xl=#3%
  \yl=#4%
  \multiply \x by \expansao%
  \multiply \y by \expansao%
  \multiply \xl by \expansao%
  \multiply \yl by \expansao%
  \!qjoin (\x,\y) (\xl,\yl)             
  \!ifnextchar){\!fim}{\!modifQjoin}}
\def\!fim){\ignorespaces}

\def\setaxy(#1{\!pontosxy #1}
\def\!pontosxy(#1,#2){%
  \!maispontosxy}
\def\!maispontosxy(#1,#2)(#3,#4){%
  \!ifnextchar){\!fimxy#3,#4}{\!maispontosxy}}
\def\!fimxy#1,#2){\x=#1%
  \y=#2
  \multiply \x by \expansao
  \multiply \y by \expansao
  \xl=\x%
  \yl=\y%
  \aux=1%
  \multiply \aux by \auxa%
  \advance\xl by \aux%
  \aux=1%
  \multiply \aux by \auxb%
  \advance\yl by \aux%
  \arrow <4pt> [.2,1.1] from {\x} {\y} to {\xl} {\yl}}

\def\cmor#1 #2(#3,#4)#5{%
  \!ifnextchar[{\!cmora{#1}{#2}{#3}{#4}{#5}}{\!cmora{#1}{#2}{#3}{#4}{#5}[0] }}
\def\!cmora#1#2#3#4#5[#6]{%
  \ifcase #2
      \auxa=0
      \auxb=1
    \or
      \auxa=0
      \auxb=-1
    \or
      \auxa=1
      \auxb=0
    \or
      \auxa=-1
      \auxb=0
    \fi  
  \ifcase #6  
    \modifplot#1
    \setaxy#1
  \or  
    \setdashes
    \modifplot#1
    \setaxy#1
    \setsolid
  \or  
    \modifplot#1
  \fi  
  \x=#3%
  \y=#4%
  \multiply \x by \expansao%
  \multiply \y by \expansao%
  \put {#5} at {\x} {\y}}

\def\obj(#1,#2){%
  \!ifnextchar[{\!obja{#1}{#2}}{\!obja{#1}{#2}[Nulo]}}
\def\!obja#1#2[#3]#4{%
  \!ifnextchar[{\!objb{#1}{#2}{#3}{#4}}{\!objb{#1}{#2}{#3}{#4}[1]}}
\def\!objb#1#2#3#4[#5]{%
  \x=#1%
  \y=#2%
  \def\!pinta{\normalsize$\bullet$}
  \def\!nulo{Nulo}%
  \def\!arg{#3}%
  \!compara{\!arg}{\!nulo}%
  \ifcompara\def\!arg{#4}\fi%
  \multiply \x by \expansao%
  \multiply \y by \expansao%
  \setbox\caixa=\hbox{#4}%
  \!coloca{(\!arg)(#1,#2)(\number\ht\caixa,\number\wd\caixa,\number\dp\caixa)}{\pilha}%
  \auxa=\wd\caixa \divide \auxa by 131072 
  \advance \auxa by 5
  \auxb=\ht\caixa
  \advance \auxb by \number\dp\caixa
  \divide \auxb by 131072 
  \advance \auxb by 5
  \ifcase \tipografo    
    \put{#4} at {\x} {\y}
  \or                   
    \ifcase #5 
      \put{#4} at {\x} {\y}
    \or        
      \put{\!pinta} at {\x} {\y}
      \advance \y by \number\auxb  
      \put{#4} at {\x} {\y}
    \or        
      \put{\!pinta} at {\x} {\y}
      \advance \auxa by -2  
      \advance \auxb by -2  
      \advance \x by \number\auxa  
      \advance \y by \number\auxb  
      \put{#4} at {\x} {\y}   
    \or        
      \put{\!pinta} at {\x} {\y}
      \advance \x by \number\auxa  
      \put{#4} at {\x} {\y}   
    \or        
      \put{\!pinta} at {\x} {\y}
      \advance \auxa by -2  
      \advance \auxb by -2  
      \advance \x by \number\auxa  
      \advance \y by -\number\auxb  
      \put{#4} at {\x} {\y}   
    \or        
      \put{\!pinta} at {\x} {\y}
      \advance \y by -\number\auxb  
      \put{#4} at {\x} {\y}   
    \or        
      \put{\!pinta} at {\x} {\y}
      \advance \auxa by -2  
      \advance \auxb by -2  
      \advance \x by -\number\auxa  
      \advance \y by -\number\auxb  
      \put{#4} at {\x} {\y}   
    \or        
      \put{\!pinta} at {\x} {\y}
      \advance \x by -\number\auxa  
      \put{#4} at {\x} {\y}   
    \or        
      \put{\!pinta} at {\x} {\y}
      \advance \auxa by -2  
      \advance \auxb by -2  
      \advance \x by -\number\auxa  
      \advance \y by \number\auxb  
      \put{#4} at {\x} {\y}   
    \fi
  \or                   
    \ifcase #5 
      \put{#4} at {\x} {\y}
    \or        
      \put{\!pinta} at {\x} {\y}
      \advance \y by \number\auxb  
      \put{#4} at {\x} {\y}
    \or        
      \put{\!pinta} at {\x} {\y}
      \advance \auxa by -2  
      \advance \auxb by -2  
      \advance \x by \number\auxa  
      \advance \y by \number\auxb  
      \put{#4} at {\x} {\y}   
    \or        
      \put{\!pinta} at {\x} {\y}
      \advance \x by \number\auxa  
      \put{#4} at {\x} {\y}   
    \or        
      \put{\!pinta} at {\x} {\y}
      \advance \auxa by -2  
      \advance \auxb by -2
      \advance \x by \number\auxa  
      \advance \y by -\number\auxb 
      \put{#4} at {\x} {\y}   
    \or        
      \put{\!pinta} at {\x} {\y}
      \advance \y by -\number\auxb 
      \put{#4} at {\x} {\y}   
    \or        
      \put{\!pinta} at {\x} {\y}
      \advance \auxa by -2  
      \advance \auxb by -2
      \advance \x by -\number\auxa 
      \advance \y by -\number\auxb 
      \put{#4} at {\x} {\y}   
    \or        
      \put{\!pinta} at {\x} {\y}
      \advance \x by -\number\auxa 
      \put{#4} at {\x} {\y}   
    \or        
      \put{\!pinta} at {\x} {\y}
      \advance \auxa by -2  
      \advance \auxb by -2
      \advance \x by -\number\auxa 
      \advance \y by \number\auxb  
      \put{#4} at {\x} {\y}   
    \fi
   \else 
     \ifnum\auxa<\auxb 
       \aux=\auxb
     \else
       \aux=\auxa
     \fi
     \ifdim\wd\caixa<1em
       \dimen99 = 1em
       \aux=\dimen99 \divide \aux by 131072 
       \advance \aux by 5
     \fi
     \advance\aux by -2 
     \multiply\aux by 2 %
     \ifnum\aux<30
       \put{\circle{\aux}} [Bl] at {\x} {\y}
     \else
       \multiply\auxa by 2
       \multiply\auxb by 2
       \put{\oval(\auxa,\auxb)} [Bl] at {\x} {\y}
     \fi
     \put{#4} at {\x} {\y}
   \fi   
}

\catcode`!=12 

  \input miniltx
  \def\Gin@driver{pdftex.def}
  \input color.sty
  \input graphicx.sty
  \resetatcatcode
%
%

%
%
%
%

\def\Serif{cmr}
\def\SerifBold{cmbx}
\def\SerifItalics{cmti}
\def\SerifSlanted{cmsl}
\def\SerifBoldItalics{cmbxti}
\def\SansSerif{cmss}
\def\SansSerifBold{cmssbx}
\def\SansSerifItalics{cmssi}
\def\SansSerifSlanted{cmssi}
\def\Math{cmmi}
\def\Symbols{cmsy}
\def\MathBold{cmmib}
\def\MoreSymbols{cmex}
\def\Typewriter{cmtt}
\def\Gothic{eufm}
\def\Double{msbm}

= 			\Serif10 			at 5pt
= 		\SerifBold10 		at 5pt
= 	\SerifItalics10 	at 5pt
=		\SerifSlanted10 	at 5pt
=	\SerifBoldItalics10	at 5pt
= 		\SansSerif10 		at 5pt
=	\SansSerifBold10	at 5pt
=	\SansSerifItalics10	at 5pt
=	\SansSerifSlanted10	at 5pt
=				\Math10				at 5pt
=			\MathBold10			at 5pt
=			\Symbols10			at 5pt
=		\MoreSymbols10		at 5pt
=		\Typewriter10		at 5pt
=			\Gothic10			at 5pt
=			\Double10			at 5pt

= 			\Serif10 			at 7pt
= 		\SerifBold10 		at 7pt
= 	\SerifItalics10 	at 7pt
=	\SerifSlanted10 	at 7pt
=\SerifBoldItalics10	at 7pt
= 		\SansSerif10 		at 7pt
= 	\SansSerifBold10 	at 7pt
=\SansSerifItalics10	at 7pt
=\SansSerifSlanted10	at 7pt
=			\Math10				at 7pt
=		\MathBold10			at 7pt
=			\Symbols10			at 7pt
=		\MoreSymbols10		at 7pt
=		\Typewriter10		at 7pt
=			\Gothic10			at 7pt
=			\Double10			at 7pt

= 			\Serif10 			at 8pt
= 		\SerifBold10 		at 8pt
= 	\SerifItalics10 	at 8pt
=	\SerifSlanted10 	at 8pt
=\SerifBoldItalics10	at 8pt
= 		\SansSerif10 		at 8pt
= 	\SansSerifBold10 	at 8pt
=\SansSerifItalics10 at 8pt
=\SansSerifSlanted10 at 8pt
=			\Math10				at 8pt
=		\MathBold10			at 8pt
=			\Symbols10			at 8pt
=		\MoreSymbols10		at 8pt
=		\Typewriter10		at 8pt
=			\Gothic10			at 8pt
=			\Double10			at 8pt

= 			\Serif10 			at 10pt
= 		\SerifBold10 		at 10pt
= 		\SerifItalics10 	at 10pt
=		\SerifSlanted10 	at 10pt
=	\SerifBoldItalics10	at 10pt
= 		\SansSerif10 		at 10pt
= 	\SansSerifBold10 	at 10pt
= 	\SansSerifItalics10 at 10pt
= 	\SansSerifSlanted10 at 10pt
=				\Math10				at 10pt
=			\MathBold10			at 10pt
=			\Symbols10			at 10pt
=		\MoreSymbols10		at 10pt
=		\Typewriter10		at 10pt
=			\Gothic10			at 10pt
=			\Double10			at 10pt

= 				\Serif10 			at 12pt
= 			\SerifBold10 		at 12pt
= 		\SerifItalics10 	at 12pt
=		\SerifSlanted10 	at 12pt
=	\SerifBoldItalics10	at 12pt
= 			\SansSerif10 		at 12pt
= 		\SansSerifBold10 	at 12pt
= 	\SansSerifItalics10 at 12pt
= 	\SansSerifSlanted10 at 12pt
=				\Math10				at 12pt
=			\MathBold10			at 12pt
=			\Symbols10			at 12pt
=		\MoreSymbols10		at 12pt
=			\Typewriter10		at 12pt
=				\Gothic10			at 12pt
=				\Double10			at 12pt

= 			\Serif10 			at 14pt
= 		\SerifBold10 		at 14pt
= 	\SerifItalics10 	at 14pt
=		\SerifSlanted10 	at 14pt
=	\SerifBoldItalics10	at 14pt
= 		\SansSerif10 		at 14pt
= 	\SansSerifBold10 	at 14pt
= \SansSerifSlanted10 at 14pt
= \SansSerifItalics10 at 14pt
=				\Math10				at 14pt
=			\MathBold10			at 14pt
=			\Symbols10			at 14pt
=		\MoreSymbols10		at 14pt
=		\Typewriter10		at 14pt
=			\Gothic10			at 14pt
=			\Double10			at 14pt

\def\NormalStyle{\parindent=5pt\parskip=3pt\normalbaselineskip=14pt%
\def\nt{\tenSerif}%
\def\rm{\fam0\tenSerif}%
\textfont0=\tenSerif\scriptfont0=\sevenSerif\scriptscriptfont0=\fiveSerif
\textfont1=\tenMath\scriptfont1=\sevenMath\scriptscriptfont1=\fiveMath
\textfont2=\tenSymbols\scriptfont2=\sevenSymbols\scriptscriptfont2=\fiveSymbols
\textfont3=\tenMoreSymbols\scriptfont3=\sevenMoreSymbols\scriptscriptfont3=\fiveMoreSymbols
\textfont\itfam=\tenSerifItalics\def\it{\fam\itfam\tenSerifItalics}%
\textfont\slfam=\tenSerifSlanted\def\sl{\fam\slfam\tenSerifSlanted}%
\textfont\ttfam=\tenTypewriter\def\tt{\fam\ttfam\tenTypewriter}%
\textfont\bffam=\tenSerifBold%
\def\bf{\fam\bffam\tenSerifBold}\scriptfont\bffam=\sevenSerifBold\scriptscriptfont\bffam=\fiveSerifBold%
\def\cal{\tenSymbols}%
\def\greekbold{\tenMathBold}%
\def\gothic{\tenGothic}%
\def\Bbb{\tenDouble}%
\def\LieFont{\tenSerifItalics}%
\nt\normalbaselines\baselineskip=14pt%
}

\def\TitleStyle{\parindent=0pt\parskip=0pt\normalbaselineskip=15pt%
\def\nt{\fourteenSansSerifBold}%
\def\rm{\fam0\fourteenSansSerifBold}%
\textfont0=\fourteenSansSerifBold\scriptfont0=\tenSansSerifBold\scriptscriptfont0=\eightSansSerifBold
\textfont1=\fourteenMath\scriptfont1=\tenMath\scriptscriptfont1=\eightMath
\textfont2=\fourteenSymbols\scriptfont2=\tenSymbols\scriptscriptfont2=\eightSymbols
\textfont3=\fourteenMoreSymbols\scriptfont3=\tenMoreSymbols\scriptscriptfont3=\eightMoreSymbols
\textfont\itfam=\fourteenSansSerifItalics\def\it{\fam\itfam\fourteenSansSerifItalics}%
\textfont\slfam=\fourteenSansSerifSlanted\def\sl{\fam\slfam\fourteenSerifSansSlanted}%
\textfont\ttfam=\fourteenTypewriter\def\tt{\fam\ttfam\fourteenTypewriter}%
\textfont\bffam=\fourteenSansSerif%
\def\bf{\fam\bffam\fourteenSansSerif}\scriptfont\bffam=\tenSansSerif\scriptscriptfont\bffam=\eightSansSerif%
\def\cal{\fourteenSymbols}%
\def\greekbold{\fourteenMathBold}%
\def\gothic{\fourteenGothic}%
\def\Bbb{\fourteenDouble}%
\def\LieFont{\fourteenSerifItalics}%
\nt\normalbaselines\baselineskip=15pt%
}

\def\PartStyle{\parindent=0pt\parskip=0pt\normalbaselineskip=15pt%
\def\nt{\fourteenSansSerifBold}%
\def\rm{\fam0\fourteenSansSerifBold}%
\textfont0=\fourteenSansSerifBold\scriptfont0=\tenSansSerifBold\scriptscriptfont0=\eightSansSerifBold
\textfont1=\fourteenMath\scriptfont1=\tenMath\scriptscriptfont1=\eightMath
\textfont2=\fourteenSymbols\scriptfont2=\tenSymbols\scriptscriptfont2=\eightSymbols
\textfont3=\fourteenMoreSymbols\scriptfont3=\tenMoreSymbols\scriptscriptfont3=\eightMoreSymbols
\textfont\itfam=\fourteenSansSerifItalics\def\it{\fam\itfam\fourteenSansSerifItalics}%
\textfont\slfam=\fourteenSansSerifSlanted\def\sl{\fam\slfam\fourteenSerifSansSlanted}%
\textfont\ttfam=\fourteenTypewriter\def\tt{\fam\ttfam\fourteenTypewriter}%
\textfont\bffam=\fourteenSansSerif%
\def\bf{\fam\bffam\fourteenSansSerif}\scriptfont\bffam=\tenSansSerif\scriptscriptfont\bffam=\eightSansSerif%
\def\cal{\fourteenSymbols}%
\def\greekbold{\fourteenMathBold}%
\def\gothic{\fourteenGothic}%
\def\Bbb{\fourteenDouble}%
\def\LieFont{\fourteenSerifItalics}%
\nt\normalbaselines\baselineskip=15pt%
}

\def\ChapterStyle{\parindent=0pt\parskip=0pt\normalbaselineskip=15pt%
\def\nt{\fourteenSansSerifBold}%
\def\rm{\fam0\fourteenSansSerifBold}%
\textfont0=\fourteenSansSerifBold\scriptfont0=\tenSansSerifBold\scriptscriptfont0=\eightSansSerifBold
\textfont1=\fourteenMath\scriptfont1=\tenMath\scriptscriptfont1=\eightMath
\textfont2=\fourteenSymbols\scriptfont2=\tenSymbols\scriptscriptfont2=\eightSymbols
\textfont3=\fourteenMoreSymbols\scriptfont3=\tenMoreSymbols\scriptscriptfont3=\eightMoreSymbols
\textfont\itfam=\fourteenSansSerifItalics\def\it{\fam\itfam\fourteenSansSerifItalics}%
\textfont\slfam=\fourteenSansSerifSlanted\def\sl{\fam\slfam\fourteenSerifSansSlanted}%
\textfont\ttfam=\fourteenTypewriter\def\tt{\fam\ttfam\fourteenTypewriter}%
\textfont\bffam=\fourteenSansSerif%
\def\bf{\fam\bffam\fourteenSansSerif}\scriptfont\bffam=\tenSansSerif\scriptscriptfont\bffam=\eightSansSerif%
\def\cal{\fourteenSymbols}%
\def\greekbold{\fourteenMathBold}%
\def\gothic{\fourteenGothic}%
\def\Bbb{\fourteenDouble}%
\def\LieFont{\fourteenSerifItalics}%
\nt\normalbaselines\baselineskip=15pt%
}

\def\SectionStyle{\parindent=0pt\parskip=0pt\normalbaselineskip=13pt%
\def\nt{\twelveSansSerifBold}%
\def\rm{\fam0\twelveSansSerifBold}%
\textfont0=\twelveSansSerifBold\scriptfont0=\eightSansSerifBold\scriptscriptfont0=\eightSansSerifBold
\textfont1=\twelveMath\scriptfont1=\eightMath\scriptscriptfont1=\eightMath
\textfont2=\twelveSymbols\scriptfont2=\eightSymbols\scriptscriptfont2=\eightSymbols
\textfont3=\twelveMoreSymbols\scriptfont3=\eightMoreSymbols\scriptscriptfont3=\eightMoreSymbols
\textfont\itfam=\twelveSansSerifItalics\def\it{\fam\itfam\twelveSansSerifItalics}%
\textfont\slfam=\twelveSansSerifSlanted\def\sl{\fam\slfam\twelveSerifSansSlanted}%
\textfont\ttfam=\twelveTypewriter\def\tt{\fam\ttfam\twelveTypewriter}%
\textfont\bffam=\twelveSansSerif%
\def\bf{\fam\bffam\twelveSansSerif}\scriptfont\bffam=\eightSansSerif\scriptscriptfont\bffam=\eightSansSerif%
\def\cal{\twelveSymbols}%
\def\bg{\twelveMathBold}%
\def\gothic{\twelveGothic}%
\def\Bbb{\twelveDouble}%
\def\LieFont{\twelveSerifItalics}%
\nt\normalbaselines\baselineskip=13pt%
}

\def\SubSectionStyle{\parindent=0pt\parskip=0pt\normalbaselineskip=13pt%
\def\nt{\twelveSansSerifItalics}%
\def\rm{\fam0\twelveSansSerifItalics}%
\textfont0=\twelveSansSerifItalics\scriptfont0=\eightSansSerifItalics\scriptscriptfont0=\eightSansSerifItalics%
\textfont1=\twelveMath\scriptfont1=\eightMath\scriptscriptfont1=\eightMath%
\textfont2=\twelveSymbols\scriptfont2=\eightSymbols\scriptscriptfont2=\eightSymbols%
\textfont3=\twelveMoreSymbols\scriptfont3=\eightMoreSymbols\scriptscriptfont3=\eightMoreSymbols%
\textfont\itfam=\twelveSansSerif\def\it{\fam\itfam\twelveSansSerif}%
\textfont\slfam=\twelveSansSerifSlanted\def\sl{\fam\slfam\twelveSerifSansSlanted}%
\textfont\ttfam=\twelveTypewriter\def\tt{\fam\ttfam\twelveTypewriter}%
\textfont\bffam=\twelveSansSerifBold%
\def\bf{\fam\bffam\twelveSansSerifBold}\scriptfont\bffam=\eightSansSerifBold\scriptscriptfont\bffam=\eightSansSerifBold%
\def\cal{\twelveSymbols}%
\def\greekbold{\twelveMathBold}%
\def\gothic{\twelveGothic}%
\def\Bbb{\twelveDouble}%
\def\LieFont{\twelveSerifItalics}%
\nt\normalbaselines\baselineskip=13pt%
}

\def\AuthorStyle{\parindent=0pt\parskip=0pt\normalbaselineskip=14pt%
\def\nt{\tenSerif}%
\def\rm{\fam0\tenSerif}%
\textfont0=\tenSerif\scriptfont0=\sevenSerif\scriptscriptfont0=\fiveSerif
\textfont1=\tenMath\scriptfont1=\sevenMath\scriptscriptfont1=\fiveMath
\textfont2=\tenSymbols\scriptfont2=\sevenSymbols\scriptscriptfont2=\fiveSymbols
\textfont3=\tenMoreSymbols\scriptfont3=\sevenMoreSymbols\scriptscriptfont3=\fiveMoreSymbols
\textfont\itfam=\tenSerifItalics\def\it{\fam\itfam\tenSerifItalics}%
\textfont\slfam=\tenSerifSlanted\def\sl{\fam\slfam\tenSerifSlanted}%
\textfont\ttfam=\tenTypewriter\def\tt{\fam\ttfam\tenTypewriter}%
\textfont\bffam=\tenSerifBold%
\def\bf{\fam\bffam\tenSerifBold}\scriptfont\bffam=\sevenSerifBold\scriptscriptfont\bffam=\fiveSerifBold%
\def\cal{\tenSymbols}%
\def\greekbold{\tenMathBold}%
\def\gothic{\tenGothic}%
\def\Bbb{\tenDouble}%
\def\LieFont{\tenSerifItalics}%
\nt\normalbaselines\baselineskip=14pt%
}

\def\AddressStyle{\parindent=0pt\parskip=0pt\normalbaselineskip=14pt%
\def\nt{\eightSerif}%
\def\rm{\fam0\eightSerif}%
\textfont0=\eightSerif\scriptfont0=\sevenSerif\scriptscriptfont0=\fiveSerif
\textfont1=\eightMath\scriptfont1=\sevenMath\scriptscriptfont1=\fiveMath
\textfont2=\eightSymbols\scriptfont2=\sevenSymbols\scriptscriptfont2=\fiveSymbols
\textfont3=\eightMoreSymbols\scriptfont3=\sevenMoreSymbols\scriptscriptfont3=\fiveMoreSymbols
\textfont\itfam=\eightSerifItalics\def\it{\fam\itfam\eightSerifItalics}%
\textfont\slfam=\eightSerifSlanted\def\sl{\fam\slfam\eightSerifSlanted}%
\textfont\ttfam=\eightTypewriter\def\tt{\fam\ttfam\eightTypewriter}%
\textfont\bffam=\eightSerifBold%
\def\bf{\fam\bffam\eightSerifBold}\scriptfont\bffam=\sevenSerifBold\scriptscriptfont\bffam=\fiveSerifBold%
\def\cal{\eightSymbols}%
\def\greekbold{\eightMathBold}%
\def\gothic{\eightGothic}%
\def\Bbb{\eightDouble}%
\def\LieFont{\eightSerifItalics}%
\nt\normalbaselines\baselineskip=14pt%
}

\def\AbstractStyle{\parindent=0pt\parskip=0pt\normalbaselineskip=12pt%
\def\nt{\eightSerif}%
\def\rm{\fam0\eightSerif}%
\textfont0=\eightSerif\scriptfont0=\sevenSerif\scriptscriptfont0=\fiveSerif
\textfont1=\eightMath\scriptfont1=\sevenMath\scriptscriptfont1=\fiveMath
\textfont2=\eightSymbols\scriptfont2=\sevenSymbols\scriptscriptfont2=\fiveSymbols
\textfont3=\eightMoreSymbols\scriptfont3=\sevenMoreSymbols\scriptscriptfont3=\fiveMoreSymbols
\textfont\itfam=\eightSerifItalics\def\it{\fam\itfam\eightSerifItalics}%
\textfont\slfam=\eightSerifSlanted\def\sl{\fam\slfam\eightSerifSlanted}%
\textfont\ttfam=\eightTypewriter\def\tt{\fam\ttfam\eightTypewriter}%
\textfont\bffam=\eightSerifBold%
\def\bf{\fam\bffam\eightSerifBold}\scriptfont\bffam=\sevenSerifBold\scriptscriptfont\bffam=\fiveSerifBold%
\def\cal{\eightSymbols}%
\def\greekbold{\eightMathBold}%
\def\gothic{\eightGothic}%
\def\Bbb{\eightDouble}%
\def\LieFont{\eightSerifItalics}%
\nt\normalbaselines\baselineskip=12pt%
}

\def\RefsStyle{\parindent=0pt\parskip=0pt%
\def\nt{\eightSerif}%
\def\rm{\fam0\eightSerif}%
\textfont0=\eightSerif\scriptfont0=\sevenSerif\scriptscriptfont0=\fiveSerif
\textfont1=\eightMath\scriptfont1=\sevenMath\scriptscriptfont1=\fiveMath
\textfont2=\eightSymbols\scriptfont2=\sevenSymbols\scriptscriptfont2=\fiveSymbols
\textfont3=\eightMoreSymbols\scriptfont3=\sevenMoreSymbols\scriptscriptfont3=\fiveMoreSymbols
\textfont\itfam=\eightSerifItalics\def\it{\fam\itfam\eightSerifItalics}%
\textfont\slfam=\eightSerifSlanted\def\sl{\fam\slfam\eightSerifSlanted}%
\textfont\ttfam=\eightTypewriter\def\tt{\fam\ttfam\eightTypewriter}%
\textfont\bffam=\eightSerifBold%
\def\bf{\fam\bffam\eightSerifBold}\scriptfont\bffam=\sevenSerifBold\scriptscriptfont\bffam=\fiveSerifBold%
\def\cal{\eightSymbols}%
\def\greekbold{\eightMathBold}%
\def\gothic{\eightGothic}%
\def\Bbb{\eightDouble}%
\def\LieFont{\eightSerifItalics}%
\nt\normalbaselines\baselineskip=10pt%
}

\def\ClaimStyle{\parindent=5pt\parskip=3pt\normalbaselineskip=14pt%
\def\nt{\tenSerifSlanted}%
\def\rm{\fam0\tenSerifSlanted}%
\textfont0=\tenSerifSlanted\scriptfont0=\sevenSerifSlanted\scriptscriptfont0=\fiveSerifSlanted
\textfont1=\tenMath\scriptfont1=\sevenMath\scriptscriptfont1=\fiveMath
\textfont2=\tenSymbols\scriptfont2=\sevenSymbols\scriptscriptfont2=\fiveSymbols
\textfont3=\tenMoreSymbols\scriptfont3=\sevenMoreSymbols\scriptscriptfont3=\fiveMoreSymbols
\textfont\itfam=\tenSerifItalics\def\it{\fam\itfam\tenSerifItalics}%
\textfont\slfam=\tenSerif\def\sl{\fam\slfam\tenSerif}%
\textfont\ttfam=\tenTypewriter\def\tt{\fam\ttfam\tenTypewriter}%
\textfont\bffam=\tenSerifBold%
\def\bf{\fam\bffam\tenSerifBold}\scriptfont\bffam=\sevenSerifBold\scriptscriptfont\bffam=\fiveSerifBold%
\def\cal{\tenSymbols}%
\def\greekbold{\tenMathBold}%
\def\gothic{\tenGothic}%
\def\Bbb{\tenDouble}%
\def\LieFont{\tenSerifItalics}%
\nt\normalbaselines\baselineskip=14pt%
}

\def\ProofStyle{\parindent=5pt\parskip=3pt\normalbaselineskip=14pt%
\def\nt{\tenSerifSlanted}%
\def\rm{\fam0\tenSerifSlanted}%
\textfont0=\tenSerif\scriptfont0=\sevenSerif\scriptscriptfont0=\fiveSerif
\textfont1=\tenMath\scriptfont1=\sevenMath\scriptscriptfont1=\fiveMath
\textfont2=\tenSymbols\scriptfont2=\sevenSymbols\scriptscriptfont2=\fiveSymbols
\textfont3=\tenMoreSymbols\scriptfont3=\sevenMoreSymbols\scriptscriptfont3=\fiveMoreSymbols
\textfont\itfam=\tenSerifItalics\def\it{\fam\itfam\tenSerifItalics}%
\textfont\slfam=\tenSerif\def\sl{\fam\slfam\tenSerif}%
\textfont\ttfam=\tenTypewriter\def\tt{\fam\ttfam\tenTypewriter}%
\textfont\bffam=\tenSerifBold%
\def\bf{\fam\bffam\tenSerifBold}\scriptfont\bffam=\sevenSerifBold\scriptscriptfont\bffam=\fiveSerifBold%
\def\cal{\tenSymbols}%
\def\greekbold{\tenMathBold}%
\def\gothic{\tenGothic}%
\def\Bbb{\tenDouble}%
\def\LieFont{\tenSerifItalics}%
\nt\normalbaselines\baselineskip=14pt%
}

%
%


\def\ModeYes{yes}
\def\ModeNo{no}

\def\ModeUndef{undefined}


\def\nx{\noexpand}
\def\ni{\noindent}
\def\newpage{\vfill\eject}

\def\ss{\vskip 5pt}
\def\ms{\vskip 10pt}
\def\bs{\vskip 20pt}

 \def\,{\mskip\thinmuskip}
 \def\!{\mskip-\thinmuskip}
 \def\>{\mskip\medmuskip}
 \def\;{\mskip\thickmuskip}

%
% ... 
%

\def\refsModePost{post}
\def\refsModeAuto{auto}

\def\dbRefsSatusModeOk{ok}
\def\dbRefsSatusModeError{error}
\def\dbRefsSatusModeWarning{warning}


\newcount\BNUM
\BNUM=0

\def\refs{}

\def\SetModePost{\xdef\refsMode{\refsModePost}}			
\SetModePost

\def\dbRefsStatusOk{%
	\xdef\dbRefsStatus{\dbRefsSatusModeOk}%
	\xdef\dbRefsError{\ModeNo}%
	\xdef\dbRefsWarning{\ModeNo}%
	\xdef\dbRefsInfo{\ModeNo}%
}

\def\dbRefs{%
}

\def\dbRefsGet#1{%
	\xdef\found{N}\xdef\ikey{#1}\dbRefsStatusOk%
	\xdef\key{\ModeUndef}\xdef\tag{\ModeUndef}\xdef\tail{\ModeUndef}%
	\dbRefs%
}

\def\NextRefsTag{%
	\global\advance\BNUM by 1%
}
\def\ShowTag#1{{\bf [#1]}}

\def\dbRefsInsert#1#2{%
\dbRefsGet{#1}%
\if\found Y %
   \xdef\dbRefsStatus{\dbRefsSatusModeWarning}%
   \xdef\dbRefsWarning{record is already there}%
   \xdef\dbRefsInfo{record not inserted}%
\else%
   \toks2=\expandafter{\dbRefs}%
   \ifx\refsMode\refsModeAuto \NextRefsTag
    \xdef\dbRefs{%
   	\the\toks2 \nx\xdef\nx\dbx{#1}%
	\nx\ifx\nx\ikey %
		\nx\dbx\nx\xdef\nx\found{Y}%
		\nx\xdef\nx\key{#1}%
		\nx\xdef\nx\tag{\the\BNUM}%
		\nx\xdef\nx\tail{#2}%
	\nx\fi}%
	\global\xdef\refs{\refs \ss\ni[\the\BNUM]\ #2\par}
   \fi%
   \ifx\refsMode\refsModePost 
    \xdef\dbRefs{%
   	\the\toks2 \nx\xdef\nx\dbx{#1}%
	\nx\ifx\nx\ikey %
		\nx\dbx\nx\xdef\nx\found{Y}%
		\nx\xdef\nx\key{#1}%
		\nx\xdef\nx\tag{\ModeUndef}%
		\nx\xdef\nx\tail{#2}%
	\nx\fi}%
   \fi%
\fi%
}

\def\dbRefsEdit#1#2#3{\dbRefsGet{#1}%
\if\found N 
   \xdef\dbRefsStatus{\dbRefsSatusModeError}%
   \xdef\dbRefsError{record is not there}%
   \xdef\dbRefsInfo{record not edited}%
\else%
   \toks2=\expandafter{\dbRefs}%
   \xdef\dbRefs{\the\toks2%
   \nx\xdef\nx\dbx{#1}%
   \nx\ifx\nx\ikey\nx\dbx %
	\nx\xdef\nx\found{Y}%
	\nx\xdef\nx\key{#1}%
	\nx\xdef\nx\tag{#2}%
	\nx\xdef\nx\tail{#3}%
   \nx\fi}%
\fi%
}

\def\bib#1#2{\RefsStyle\dbRefsInsert{#1}{#2}%
	\ifx\dbRefsStatus\dbRefsSatusModeWarning %
		\message{^^J}%
		\message{WARNING: Reference [#1] is doubled.^^J}%
	\fi%
}

\def\ref#1{\dbRefsGet{#1}%
\ifx\found N %
  \message{^^J}%
  \message{ERROR: Reference [#1] unknown.^^J}%
  \ShowTag{??}%
\else%
	\ifx\tag\ModeUndef \NextRefsTag%
		\dbRefsEdit{#1}{\the\BNUM}{\tail}%
		\dbRefsGet{#1}%
		\global\xdef\refs{\refs \ss\ni [\tag]\ \tail\par}
	\fi
	\ShowTag{\tag}%
\fi%
}

\def\EqRef#1{\ShowLabel{#1}}

\def\dbLabelPreInsert#1#2{\dbLabelGet{#1}%
\if\found Y %
  \xdef\dbLabelStatus{\dbLabelSatusModeWarning}%
   \xdef\dbLabelWarning{Label is already there}%
   \xdef\dbLabelInfo{Label not inserted}%
   \message{^^J}%
   \errmessage{Double pre definition of label [#1]^^J}%
\else%
   \toks2=\expandafter{\dbLabel}%
    \xdef\dbLabel{%
   	\the\toks2 \nx\xdef\nx\dbx{#1}%
	\nx\ifx\nx\ikey %
		\nx\dbx\nx\xdef\nx\found{Y}%
		\nx\xdef\nx\key{#1}%
		\nx\xdef\nx\tag{#2}%
		\nx\xdef\nx\pre{\ModeYes}%
	\nx\fi}%
\fi%
}

\def\ShowBiblio{\bs\Ensure{\SectionEnsure}%
{\SectionStyle\ni References}%
{\RefsStyle\refs}%
}

\newcount\CHANGES
\CHANGES=0
\def\AuxFile{7}
\def\PreventDoubleOn{\xdef\PreventDoubleLabel{\ModeYes}}

\PreventDoubleOn

\def\StoreLabel#1#2{\xdef\itag{#2}
 \ifx\PreModeStatus\ModeNo %
   \message{^^J}%
   \errmessage{You can't use Check without starting with OpenPreMode (and finishing with ClosePreMode)^^J}%
 \else%
   \immediate\write\AuxFile{\nx\dbLabelPreInsert{#1}{\itag}}%
   \dbLabelGet{#1}%
   \ifx\itag\tag %
   \else%
	\global\advance\CHANGES by 1%
 	\xdef\itag{(?.??)}%
    \fi%
   \fi%
}

\def\PreModeStatus{\ModeNo}

\def\ClosePreMode{\immediate\closeout\AuxFile%
  \ifnum\CHANGES=0%
	\message{^^J}%
	\message{**********************************^^J}%
	\message{**  NO CHANGES TO THE AuxFile  **^^J}%
	\message{**********************************^^J}%
 \else%
	\message{^^J}%
	\message{**************************************************^^J}%
	\message{**  PLAEASE TYPESET IT AGAIN (\the\CHANGES)  **^^J}%
    \errmessage{**************************************************^^ J}%
  \fi%
  \edef\PreModeStatus{\ModeNo}%
}

\def\dbLabelSatusModeOk{ok}

\def\dbLabelSatusModeWarning{warning}

\def\dbLabelStatusOk{%
	\xdef\dbLabelStatus{\dbLabelSatusModeOk}%
	\xdef\dbLabelError{\ModeNo}%
	\xdef\dbLabelWarning{\ModeNo}%
	\xdef\dbLabelInfo{\ModeNo}%
}

\def\dbLabel{%
}

\def\dbLabelGet#1{%
	\xdef\found{N}\xdef\ikey{#1}\dbLabelStatusOk%
	\xdef\key{\ModeUndef}\xdef\tag{\ModeUndef}\xdef\pre{\ModeUndef}%
	\dbLabel%
}

\def\ShowLabel#1{%
 \dbLabelGet{#1}%
 \ifx\tag \ModeUndef %
 	\global\advance\CHANGES by 1%
 	(?.??)%
 \else%
 	\tag%
 \fi%
}

\def\dbLabelPreInsert#1#2{\dbLabelGet{#1}%
\if\found Y %
  \xdef\dbLabelStatus{\dbLabelSatusModeWarning}%
   \xdef\dbLabelWarning{Label is already there}%
   \xdef\dbLabelInfo{Label not inserted}%
   \message{^^J}%
   \errmessage{Double pre definition of label [#1]^^J}%
\else%
   \toks2=\expandafter{\dbLabel}%
    \xdef\dbLabel{%
   	\the\toks2 \nx\xdef\nx\dbx{#1}%
	\nx\ifx\nx\ikey %
		\nx\dbx\nx\xdef\nx\found{Y}%
		\nx\xdef\nx\key{#1}%
		\nx\xdef\nx\tag{#2}%
		\nx\xdef\nx\pre{\ModeYes}%
	\nx\fi}%
\fi%
}

\def\dbLabelInsert#1#2{\dbLabelGet{#1}%
\xdef\itag{#2}%
\dbLabelGet{#1}%
\if\found Y %
	\ifx\tag\itag %
	\else%
	   \ifx\PreventDoubleLabel\ModeYes %
		\message{^^J}%
		\errmessage{Double definition of label [#1]^^J}%
	   \else%
		\message{^^J}%
		\message{Double definition of label [#1]^^J}%
	   \fi%
	\fi%
   \xdef\dbLabelStatus{\dbLabelSatusModeWarning}%
   \xdef\dbLabelWarning{Label is already there}%
   \xdef\dbLabelInfo{Label not inserted}%
\else%
   \toks2=\expandafter{\dbLabel}%
    \xdef\dbLabel{%
   	\the\toks2 \nx\xdef\nx\dbx{#1}%
	\nx\ifx\nx\ikey %
		\nx\dbx\nx\xdef\nx\found{Y}%
		\nx\xdef\nx\key{#1}%
		\nx\xdef\nx\tag{#2}%
		\nx\xdef\nx\pre{\ModeNo}%
	\nx\fi}%
\fi%
}


\newcount\PART
\newcount\CHAPTER
\newcount\SECTION
\newcount\SUBSECTION
\newcount\FNUMBER

\PART=0
\CHAPTER=0
\SECTION=0
\SUBSECTION=0	
\FNUMBER=0

\def\LastPart{\ModeUndef}
\def\LastChapter{\ModeUndef}
\def\LastSection{\ModeUndef}
\def\LastSubSection{\ModeUndef}
\def\LastClaim{\ModeUndef}
\def\Last{\ModeUndef}

\newdimen\TOBOTTOM
\newdimen\LIMIT

\def\Ensure#1{\ \par\ \immediate\LIMIT=#1\immediate\TOBOTTOM=\the\pagegoal\advance\TOBOTTOM by -\pagetotal%
\ifdim\TOBOTTOM<\LIMIT\newpage \else%
\vskip-\parskip\vskip-\parskip\vskip-\baselineskip\fi}

\def\PartLabel{\the\PART}
\def\NewPart#1{\global\advance\PART by 1%
         \bs\ni{\PartStyle  Part \PartLabel:}
         \bs\ni{\PartStyle #1}\newpage%
         \CHAPTER=0\SECTION=0\SUBSECTION=0\FNUMBER=0%
         \gdef\Left{#1}%
         \global\edef\Last{\PartLabel}%
         \global\edef\LastPart{\PartLabel}%
         \global\edef\LastChapter{\ModeUndef}%
         \global\edef\LastSection{\ModeUndef}%
         \global\edef\LastSubSection{\ModeUndef}%
         \global\edef\LastClaim{\ModeUndef}}
\def\ChapterLabel{\the\CHAPTER}
\def\NewChapter#1{\global\advance\CHAPTER by 1%
         \bs\ni{\ChapterStyle  Chapter \ChapterLabel: #1}\ms%
         \SECTION=0\SUBSECTION=0\FNUMBER=0%
         \gdef\Left{#1}%
         \global\edef\Last{\ChapterLabel}%
         \global\edef\LastChapter{\ChapterLabel}%
         \global\edef\LastSection{\ModeUndef}%
         \global\edef\LastSubSection{\ModeUndef}%
         \global\edef\LastClaim{\ModeUndef}}
\def\SectionEnsure{3cm}
\def\NewSection#1{\Ensure{\SectionEnsure}\gdef\SectionLabel{\the\SECTION}\global\advance\SECTION by 1%
         \bs\ni{\SectionStyle  \SectionLabel.\ #1}\ss%
         \SUBSECTION=0\FNUMBER=0%
         \gdef\Left{#1}%
         \global\edef\Last{\SectionLabel}%
         \global\edef\LastSection{\SectionLabel}%
         \global\edef\LastSubSection{\ModeUndef}%
         \global\edef\LastClaim{\ModeUndef}}
\def\NewAppendix#1#2{\Ensure{\SectionEnsure}\gdef\SectionLabel{#1}\global\advance\SECTION by 1%
         \bs\ni{\SectionStyle  Appendix \SectionLabel.\ #2}\ss%
         \SUBSECTION=0\FNUMBER=0%
         \gdef\Left{#2}%
         \global\edef\Last{\SectionLabel}%
         \global\edef\LastSection{\SectionLabel}%
         \global\edef\LastSubSection{\ModeUndef}%
         \global\edef\LastClaim{\ModeUndef}}
\def\Acknowledgements{\Ensure{\SectionEnsure}\gdef\SectionLabel{}%
         \bs\ni{\SectionStyle  Acknowledgments}\ss%
         \SECTION=0\SUBSECTION=0\FNUMBER=0%
         \gdef\Left{}%
         \global\edef\Last{\ModeUndef}%
         \global\edef\LastSection{\ModeUndef}%
         \global\edef\LastSubSection{\ModeUndef}%
         \global\edef\LastClaim{\ModeUndef}}
\def\SubSectionEnsure{2cm}
\def\SubSectionLabel{\ifnum\SECTION>0 \the\SECTION.\fi\the\SUBSECTION}
\def\NewSubSection#1{\Ensure{\SubSectionEnsure}\global\advance\SUBSECTION by 1%
         \ms\ni{\SubSectionStyle #1}\ss%
         \global\edef\Last{\SubSectionLabel}%
         \global\edef\LastSubSection{\SubSectionLabel}}
\def\SetNumberingModeN{\def\ClaimLabel{(\the\FNUMBER)}}
\def\SetNumberingModeSN{\def\ClaimLabel{(\ifnum\SECTION>0 \SectionLabel.\fi%
      \the\FNUMBER)}}
\def\SetNumberingModeCSN{\def\ClaimLabel{(\ifnum\CHAPTER>0 \the\CHAPTER.\fi%
      \ifnum\SECTION>0 \SectionLabel.\fi%
      \the\FNUMBER)}}

\def\NewClaim{\global\advance\FNUMBER by 1%
    \ClaimLabel%
    \global\edef\LastClaim{\ClaimLabel}%
    \global\edef\Last{\ClaimLabel}}

\def\HideLabels{\xdef\ShowLabelsMode{\ModeNo}}
\HideLabels

\def\fn{\eqno{\NewClaim}} 
\def\fl#1{%
\ifx\ShowLabelsMode\ModeYes%
 \eqno{{\buildrel{\hbox{\AbstractStyle[#1]}}\over{\hfill\NewClaim}}}%
\else%
 \eqno{\NewClaim}%
\fi%
\dbLabelInsert{#1}{\ClaimLabel}}
\def\fprel#1{\global\advance\FNUMBER by 1\StoreLabel{#1}{\ClaimLabel}%
\ifx\ShowLabelsMode\ModeYes%
\eqno{{\buildrel{\hbox{\AbstractStyle[#1]}}\over{\hfill.\itag}}}%
\else%
 \eqno{\itag}%
\fi%
}

\def\cl#1{\global\advance\FNUMBER by 1\dbLabelInsert{#1}{\ClaimLabel}%
\ifx\ShowLabelsMode\ModeYes%
${\buildrel{\hbox{\AbstractStyle[#1]}}\over{\hfill\ClaimLabel}}$%
\else%
  $\ClaimLabel$%
\fi%
}
\def\cprel#1{\global\advance\FNUMBER by 1\StoreLabel{#1}{\ClaimLabel}%
\ifx\ShowLabelsMode\ModeYes%
${\buildrel{\hbox{\AbstractStyle[#1]}}\over{\hfill.\itag}}$%
\else%
  $\itag$%
\fi%
}


\parindent=7pt
\leftskip=2cm
\newcount\SideIndent
\newcount\SideIndentTemp
\SideIndent=0
\newdimen\SectionIndent
\SectionIndent=-8pt

\def\sidebar{\vrule height15pt width.2pt }
\def\endcorner{\hbox{\hbox{\vrule height6pt width.2pt}\vbox to6pt{\vfill\hbox
to4pt{\leaders\hrule height0.2pt\hfill}}}}
\def\begincorner{\hbox{\hbox{\vrule height6pt width.2pt}\vbox to6pt{\hbox
to4pt{\leaders\hrule height0.2pt\hfill}}}}
\def\endbegincorner{\hbox{\vbox to15pt{\endcorner\vskip-6pt\begincorner\vfill}}}
\def\SideShow{\SideIndentTemp=\SideIndent \ifnum \SideIndentTemp>0 
\loop\sidebar\hskip 2pt \advance\SideIndentTemp by-1\ifnum \SideIndentTemp>1 \repeat\fi}

\def\BeginSection{{\vbadness 100000 \par\ni\hskip\SectionIndent%
\SideShow\vbox to 15pt{\vfill\begincorner}}\global\advance\SideIndent by1\vskip-10pt}

\def\EndSection{{\vbadness 100000 \par\ni\global\advance\SideIndent by-1%
\hskip\SectionIndent\SideShow\vbox to15pt{\endcorner\vfill}\vskip-10pt}}

\def\EndBeginSection{{\vbadness 100000\par\ni%
\global\advance\SideIndent by-1\hskip\SectionIndent\SideShow
\vbox to15pt{\vfill\endbegincorner}}%
\global\advance\SideIndent by1\vskip-10pt}

\def\ShowBeginCorners#1{%
\SideIndentTemp =#1 \advance\SideIndentTemp by-1%
\ifnum \SideIndentTemp>0 %
\vskip-15truept\hbox{\kern 2truept\vbox{\hbox{\begincorner}%
\ShowBeginCorners{\SideIndentTemp}\vskip-3truept}}%
\fi%
}

\def\ShowEndCorners#1{%
\SideIndentTemp =#1 \advance\SideIndentTemp by-1%
\ifnum \SideIndentTemp>0 %
\vskip-15truept\hbox{\kern 2truept\vbox{\hbox{\endcorner}%
\ShowEndCorners{\SideIndentTemp}\vskip 2truept}}%
\fi%
}

\def\BeginSections#1{{\vbadness 100000 \par\ni\hskip\SectionIndent%
\SideShow\vbox to 15pt{\vfill\ShowBeginCorners{#1}}}\global\advance\SideIndent by#1\vskip-10pt}

\def\EndSections#1{{\vbadness 100000 \par\ni\global\advance\SideIndent by-#1%
\hskip\SectionIndent\SideShow\vbox to15pt{\vskip15pt\ShowEndCorners{#1}\vfill}\vskip-10pt}}

\def\EndBeginSections#1#2{{\vbadness 100000\par\ni%
\global\advance\SideIndent by-#1%
\hbox{\hskip\SectionIndent\SideShow\kern-2pt%
\vbox to15pt{\vskip15pt\ShowEndCorners{#1}\vskip4pt\ShowBeginCorners{#2}}}}%
\global\advance\SideIndent by#2\vskip-10pt}



\def\Note{\ms\leftskip 3cm\rightskip 1.5cm\AbstractStyle}
\def\endNote{\par\leftskip 2cm\rightskip 0cm\NormalStyle\ss}

\def\CollapseAllCNotes{\long\def\CNote##1{}}
\def\ExpandAllCNotes{\long\def\CNote##1{%
\BeginSection
	\Note%
 		##1%
	\endNote%
\EndSection%
}}
\ExpandAllCNotes


\def\frame#1{\vbox{\hrule\hbox{\vrule\vbox{\kern2pt\hbox{\kern2pt#1\kern2pt}\kern2pt}\vrule}\hrule\kern-4pt}}

\def\uline#1{\underline{#1}}

\def\Box to #1#2#3{\frame{\vtop{\hbox to #1{\hfill #2 \hfill}\hbox to #1{\hfill #3 \hfill}}}}


%
%


\def\al{\alpha}
\def\be{\beta}
\def\de{\delta}
\def\ga{\gamma}

\def\ep{\epsilon}

\def\si{\sigma}

\def\Ga{\Gamma}

\def\Si{\Sigma}

\def\GaT#1#2#3{\,{}^{3}\Ga^{#1}_{#2 #3}}


\def\Lor{{\hbox{Lor}}}

\def\id{{\hbox{\rm id}}}






 \def\R{{\hbox{\Bbb R}}}

 \def\E{{\hbox{\Bbb E}}}

 \def\R{{\hbox{\Bbb R}}}


\def\Lor{{\hbox{Lor}}}

\def\id{{\hbox{\rm id}}}

\def\ip{\hbox to4pt{\leaders\hrule height0.3pt\hfill}\vbox to8pt{\leaders\vrule width0.3pt\vfill}\kern 2pt}
\def\QDE{\hfill\hbox{\ }\vrule height4pt width4pt depth0pt} 
\def\del{\partial}
\def\na{\nabla}

\def\Lie{\hbox{\LieFont \$}}

\def\arr{\rightarrow}

\def\then{\Rightarrow}


%
%

\long\def\title#1{\centerline{\TitleStyle\ni#1}}
\long\def\moretitle#1{\centerline{\TitleStyle\ni#1}}
\long\def\author#1{\ms\centerline{\AuthorStyle by {\it #1}}}

\long\def\address#1{\ss\centerline{\AddressStyle #1}\par}
\long\def\moreaddress#1{\centerline{\AddressStyle #1}\par}
\def\abstract{\ms\leftskip 3cm\rightskip .5cm\AbstractStyle{\bf \ni Abstract:}\ }
\def\endabstract{\par\leftskip 2cm\rightskip 0cm\NormalStyle\ss}

\def\THEOREM{\ClaimStyle\ni{\bf Theorem: }}

\def\THEOREMl#1{\ClaimStyle\ni{\bf Theorem \cl{#1}: }}

\def\ENDTHEOREM{\NormalStyle}

\def\PROOF{\ProofStyle\ni{\bf Proof: }}
\def\ENDPROOF{\hfill\QDE\NormalStyle}

\def\cases#1{\left\{\eqalign{#1}\right.}
\NormalStyle
\SetNumberingModeSN
\PreventDoubleOn

\SetNumberingModeSN

\def\frac[#1/#2]{\hbox{$#1\over#2$}}
\def\Frac#1#2{{#1\over#2}}
\def\({\left(}
\def\){\right)}
\def\[{\left[}
\def\]{\right]}
\def\^#1{{}^{#1}_{\>\cdot}}
\def\_#1{{}_{#1}^{\>\cdot}}
\def\Label=#1{{\buildrel {\hbox{\fiveSerif \ShowLabel{#1}}}\over =}}
\def\<{\kern -1pt}

%
%
  \definecolor{black}{rgb}{0,0,0}
  \definecolor{darkgrey}{rgb}{0.2,0.2,0.2}
  \definecolor{grey}{rgb}{0.5,0.5,0.5}
  \definecolor{lightgrey}{rgb}{0.8,0.8,0.8}
  \definecolor{white}{rgb}{1,1,1}

  \definecolor{lightred}{rgb}{1,0.5,0.5}
  \definecolor{red}{rgb}{1,0,0}
  \definecolor{darkred}{rgb}{0.5,0,0}
  \definecolor{verylightblue}{rgb}{0.8,0.8,1}
  \definecolor{lightblue}{rgb}{0.5,0.5,1}
  \definecolor{blue}{rgb}{0,0,1}
  \definecolor{darkblue}{rgb}{0,0,0.8}
  \definecolor{lightgreen}{rgb}{0.5,1,0.5}
  \definecolor{green}{rgb}{0,1,0}
  \definecolor{darkgreen}{rgb}{0,0.8,0}

  \definecolor{lightyellow}{rgb}{1,1,0.5}
  \definecolor{yellow}{rgb}{1,1,0}
  \definecolor{darkyellow}{rgb}{0.8,0.8,0}


\newcount\ItemNum

\def\NewItem{
\itemitem{(\the\ItemNum)}
\advance\ItemNum by 1
}

%
%
\def\frame#1{\vbox{\hrule\hbox{\vrule\vbox{\kern2pt\hbox{\kern2pt#1\kern2pt}\kern2pt}\vrule}\hrule\kern-4pt}}

\def\uline#1{\underline{#1}}

\def\Box to #1#2#3{\frame{\vtop{\hbox to #1{\hfill #2 \hfill}\hbox to #1{\hfill #3 \hfill}}}}


\bib{TAY1}{M. Taylor, {\it Partial Differential Equations I}, Springer, New York (1996).}

\bib{TAY2}{M. Taylor, {\it Partial Differential Equations II}, Springer, New York (1996).}

\bib{TAY3}{M. Taylor, {\it Partial Differential Equations III}, Springer, New York (1996).}

\bib{ADM}{L. Fatibene, M. Ferraris, M. Francaviglia, L. Lusanna, {\it ADM Pseudotensors, Conserved Quantities and Covariant Conservation Laws in General Relativity}, arXiv:gr-qc/1007.4071.}

\bib{Book}{L.\ Fatibene, M.\ Francaviglia, 
{\it Natural and gauge natural formalism for classical field theories. A geometric perspective including spinors and gauge theories}, 
Kluwer Academic Publishers, Dordrecht, (2003).}

\bib{R1}{L. Fatibene, M.Francaviglia, C.Rovelli, {\it On a Covariant Formulation of the Barberi-Immirzi Connection}
CQG 24 (2007) 3055-3066; gr-qc/0702134}

\bib{R2}{L. Fatibene, M.Francaviglia, C.Rovelli, {\it Spacetime Lagrangian Formulation of Barbero-Immirzi Gravity} 
CQG 24 (2007) 4207-4217; gr-qc/0706.1899}

\bib{Smirnov}{A.L. Smirnov, (private communication)}

\bib{Samuel}{J.\ Samuel, {\it Is Barbero's Hamiltonian Formulation a Gauge Theory of Lorentzian Gravity?}, Class.\ Quantum Grav.\ {\it 17}, 2000, 141-148}

\bib{Barbero}{F.\ Barbero, {\it Real Ashtekar variables for Lorentzian signature space-time},
Phys.\ Rev.\ {\it D51}, 5507, 1996}

\bib{Immirzi}{G.\ Immirzi, {\it Quantum Gravity and Regge Calculus},
Nucl.\ Phys.\ Proc.\ Suppl.\ {\bf 57}, 65-72}

\bib{RovelliBook}{C.\ Rovelli, {\it Quantum Gravity}, Cambridge University Press, Cambridge, 2004}

\bib{Thie2006}{T. Thiemann,
{\it Loop Quantum Gravity: An Inside View}, hep-th/0608210}

\bib{Gatto}{M.Ferraris, M.Francaviglia, L.Gatto, 
{\it Reducibility of $G$-invariant Linear Connections in Principal $G$-bundles}, 
Coll. Math. Societatis J\'anos Bolyai, {\bf 56} Differential Geometry, (1989), 231-252}

\bib{GM}{M. Godina, P. Matteucci,
{\it Reductive $G$-structures and Lie derivatives},
Journal of Geometry and Physics {\bf 47} (2003) 66--86
}

\bib{HolClass}{M. Berger, 
{\it Sur les groupes dÕholonomie des vari\'et\'es \`a connexion affine et des vari\'et\'es Riemanniennes},
 Bull. Soc. Math. France {\bf 83}, 279-330 (1955)}

\bib{HolClassExc}{S. Merkulov, L.Schwachh\"ofer
{\it ClassiÞcation of irreducible holonomies of torsion-free affine connections},
Annals of Mathematics, {\bf 150}(1) (1999), 77 - 149; [arXiv: math/9907206]}

\bib{Antonsen}{F.\ Antonsen, M.S.N.\ Flagga, 
{\it Spacetime Topology (I) - Chirality and the Third Stiefel-Whitney Class}, 
Int.\ J.\ Th.\ Phys.\ {\bf 41}(2), 2002}

\bib{KobaNu}{S.\ Kobayashi, K.\ Nomizu,
  {\it Foundations of differential geometry},
  John Wiley \& Sons, Inc., New York, 1963 USA}

\bib{Uni1}{L.Fatibene, M.Ferraris, M.Francaviglia, 
{\it New Cases of Universality Theorem for Gravitational Theories}, 
Classical and Quantum Gravity {\bf 27}, 165021 (2010); arXive: 1003.1617 
}

\bib{Uni2}{L.Fatibene, M.Ferraris, M.Francaviglia, 
{\it Extended Loop Quantum Gravity},
Classical and Quantum Gravity {\bf 27}, 185016 (2010); arXiv:1003.1619}

\bib{Alexandrov}{S.Alexandrov,
{\it On choice of connection in loop quantum gravity}, 
Phys.Rev. D65 (2002) 024011}

\bib{Livine}{S.Alexandrov, E.R.Livine,
{\it $SU(2)$ loop quantum gravity seen from covariant theory}
Phys.Rev. D67 (2003) 044009}

\bib{Holst}{S. Holst, 
{\it BarberoÕs Hamiltonian Derived from a Generalized Hilbert-Palatini Action}, 
Phys. Rev. D{\bf 53}, 5966, 1996}

\bib{semPI}{N. Bodendorfer, T. Thiemann, A. Thurn,
{\it New Variables for Classical and Quantum Gravity in all Dimensions III. Quantum Theory},
arXiv:1105.3705v1 [gr-qc]}

\bib{2times}{J. A. Nieto,
{\it Canonical Gravity in Two Time and Two Space Dimensions}, 
arXiv:1107.0718v3 [gr-qc]
}

\bib{NostroBI}{L. Fatibene,  M. Francaviglia, M. Ferraris,
{\it Inducing Barbero-Immirzi connections along SU(2) reductions of bundles on spacetime},
Phys. Rev. D, {\bf 84}(6),  064035 (2011)}

\bib{Maple}{K.Chu, C.Farel, G.Fee, R.McLenaghan, Fields Inst. Comm. {\bf 15}, (1997)}

\bib{Hebey}{E. Hebey, {\it Nonlinear analysis on manifolds: Sobolev spaces and inequalities}, Courant Lect. Notes Math., Vol. 5, Courant Institute of Mathematical Sciences, New York University, New York, 1999.}

\bib{Gourgoulhon}{ E. Gourgoulhon. {\it 3+1 Formalism and Bases of Numerical Relativity}, arXiv:gr-qc/0703035,  (2007).}

\bib{Rezzolla}{L.Rezzolla, O.Zanotti, {\it Relativistic Hydrodynamics}, (2013).}

\bib{Ash}{A. Ashtekar, {\it New Canonical Gravity}, Bibliopolis, (1988)}

\bib{NG}{T.W. Baumgarte, S.L. Shapiro, {\it Numerical Gravity, Solving Einstein's Equations on the Computer}, (2010).}

\bib{Norton}{ J.D. Norton, {\it General Covariance, Gauge Theories and the Kretschmann Objection}, (2001) [Preprint].}

\bib{PhysState}{L. Fatibene, M. Ferraris, G. Magnano, Constraining the Physical State by Symmetries (in preparation).}

\bib{Gotay1}{M.J. Gotay, J. Isenberg, J.E. Marsden, R.Montgomery, {\it Momentum Maps and Classical Relativistic Fields, Part I: Covariant Field Theory}, arXiv:physics/9801019v2.}

\bib{Gotay2}{M.J. Gotay, J. Isenberg, J.E. Marsden, {\it Momentum Maps and Classical Relativistic Fields, Part II: Canonical Analysis of Field Theories}, arXiv:math-ph/0411032.}

\bib{Gotay3}{M.J. Gotay, J.E. Marsden, {\it Momentum Maps and Classical Relativistic Fields, Part III: Gauge Symmetries and Initial Value
Constraints}.}

\bib{Gotay4}{J.E. Marsden, S. Pekarsky, S. Shkoller, M. West, , {\it Variational methods, multisymplectic geometry and continuum mechanics}, Journal of Geometry and Physics, Volume 38, Issues 3-4, (2001).}

\bib{ADMOriginal}{R.  Arnowitt, S.  Deser and C.  W.  Misner, in: 
{\it Gravitation: An Introduction to Current Research}, 
L. Witten ed. Wyley,  227, (New York, 1962); gr-qc/0405109
}

\bib{RT}{T.\ Regge, C.\ Teitelboim, Annals of Physics {\bf 88}, 286  (1974)}

\bib{Nester}{J. Nester, J. Isenberg,
{\it Canonical Gravity},
in: {\it General Relativity and Gravitation. One Hundred Years After the Birth of Albert Einstein. Vol. 1}
A. Held eds.
Plenum Press NY (1980)}

\bib{Lusanna}{L.Lusanna,
{\it Canonical ADM tetrad gravity: From metrological inertial gauge variables to dynamical tidal Dirac observables},
Int. J. Geom. Met. Mod. Phys., {\bf 12}(3), 2015 1530001; arXiv:1108.3224 [gr-qc]
}

\bib{LHole}{L.Lusanna, M.Pauri,
{\it Explaining Leibniz-equivalence as difference of non-inertial appearances: dis-solution of the Hole Argument and physical individuation of point-events},
In: Studies in History and Philosophy of Science Part B: Studies in History and Philosophy of Modern Physics
{\bf 37}(4), 2006,  692-725; arXiv:gr-qc/0604087
}

\long\def\Old#1{}

\NormalStyle
\NormalStyle

\CollapseAllCNotes

\title{The Cauchy problem in General Relativity:}
\moretitle{An algebraic characterization.}

\author{L. Fatibene$^{1, 2}$, S.Garruto$^{1,2}$}

\address{$^1$ Department of Mathematics, University of Torino (Italy)}

\moreaddress{$^2$ INFN- Iniz.~Spec.~QGSKY (Italy)}

\abstract
In this paper we shall analyse the structure of the Cauchy Problem (CP briefly) for General Relativity (GR briefly)
by applying the theory of first order symmetric hyperbolic systems.
The role of harmonic coordinates will be discussed.
\endabstract

\NewSection{Introduction}

The Cauchy Problem (CP) for standard General Relativity (GR) has been studied in several papers, from numerical viewpoint (see, for example, \ref{Gourgoulhon}, \ref{NG}, \ref{Rezzolla}), as well as from an analytical viewpoint (see \ref{TAY1}, \ref{TAY2}, \ref{TAY3}).
It is at the basis of all applications of numerical gravity as well as at the basis of the physical interpretation of GR theory.

The analysis of the CP for GR is essentially based on Arnowitt, Deser, and Misner (ADM) seminal paper; see \ref{ADMOriginal} and also \ref{RT}, \ref{Nester}, \ref{ADM}.
The ADM decomposition is introduced and particular coordinate systems, called {\it harmonic coordinates}, are used to study the conditions  under which one has a well-posed CP.

In this paper we shall review such a procedure aiming to clarify the role of the choice of harmonic coordinates. 
In fact, in GR one is analysing covariant equations, the Cauchy theorem is stated in terms of the principal symbol of the differential operator which is defined intrinsically (as we shall show also in GR in which the equations are non-linear but quasi-linear and the principal symbol depends on fields) and hence it does not depend on the choice of coordinates. 
This is a first step in a project aiming to characterise integrability and constraints (as done in a Hamiltonian framework) though in a Lagrangian setting.

The ADM decompositions transforms the original (vacuum) Einstein field equations (which are 10 PDE  in $\dim(M)=4$) in a new system formed by six equations, which will be a hyperbolic system, and four constraint equations. Only the first system of 6 equations is necessary to define a CP but we have to keep in mind that the second one is important as much as the first one for the physical problem and it constrains the allowed initial conditions.

ADM decomposition also allows  us to define a parameter $\tau$ which represents the evolution parameter of the system.
In other words, from a mathematical point of view, ADM decomposition is a choice of a bundle structure $(M, \R, \tau, \Si)$ over the spacetime $M$, namely:
$$
\tau \colon M \to \R.
\fl{BundleStructure}
$$
The standard fiber $\Si$ is a model for the isochronous space submanisolds $\Si_{t_0}:= \tau^{-1}(t_0)\subset M$.

The ADM transformations are the transformations which preserve the bundle structure:
$$
\cases{
&x'^{0} = x'^{0}(x^0) \cr
&x'^{i} = x'^{i}(x^0, x^i)
}
\fl{ADMTransf}
$$
where $(x^0, x^i)$ are fibered coordinates over $M$ (see~\ref{ADM}, for further details).

On the spacetime $M$, one can restrict to Lorentzian metrics $g$ for which the fibers $\Si_{t_0}$ are space--like submanifolds.
One can then decompose the metric tensor  $g_{\mu \nu}$ (or its inverse $g^{\mu \nu}$) in the following way:
$$
g_{\mu \nu} = \left(
\matrix{
-N^{2} + \vert \vec{N} \vert^{2} & N_{i} \cr
N_{j} & \ga_{ij}
}
\right), \quad
g^{\mu \nu} = \left(
\matrix{
- N^{-2} & N^{-2}N^{j} \cr
N^{-2} N^{i} & \ga^{ij} -N^{-2} N^{i} N^{j}
}
\right)
\fl{ADM4Metric}
$$
where $N$ is a spatial scalar field called the {\it lapse}, $N^i$ a spatial vector called the {\it shift} and $\ga_{ij}$ a 3-Euclidean metric
called the {\it induced metric}, defined on $\Si$, with respect to transformations on $\Si$ (see \ref{ADM}).

Let $\vec n$ be the (future directed) unit vector $g$-orthogonal  to $\Si_{t}$ and $e_i$ a basis for  vectors tangent to the fibers $\Si_{t_0}$. Then the
lapse and shift are defined by the relation 
$$
\del_0= N \vec n + N^i e_i
\fn$$

Ricci tensor, written in these new fields and in the frame $(n, e_i)$, is:
$$
\cases{
&\uline{R}_{00} = - \Frac{1}{N}\left( \de_0 K - D_i D^i N \right) - K^{ij} K_{ij}\cr
&\uline{R}_{0j} = D_l \left( K_{j.}{}^l  - \de_j^l K \right) = \uline{R}_{j0} \cr
&\uline{R}_{ji} = \Frac{1}{N} \left( \ga_{jl} \de_0 K^l_{.\,\,i} -D_i D_j N \right) + {}^3R_{ji} 
+ K K_{ji}
}
\fl{RicciTensor}
$$
where $\de_0 = \del_0 - \Lie_{\vec N}$, $K_{ij}$ is the extrinsic curvature (namely $K_{ij} = \Frac{1}{2N} \de_0 \ga_{ij}$) and $D_i$ is the covariant derivative with respect to the affine connection ${}^3 \Ga_{ij}^l$ induced by $\ga_{ij}$.

The Ricci scalar is:
$$
R = \Frac{2}{N} \left(
\de_0 K - D_i D^i N
\right)
+ {}^3R + K^2 + K_{ij} K^{ij}
\fl{RicciScalar}
$$

Einstein equations do not determine the evolution of  $N$ and $N^i$, i.e.~they are not dynamical fields, so they can be arbitrarily chosen. We will set hereafter, for sake of the simplicity, $N = 1$ and $N^j = 0$.

With these choices we have that the metric tensor and its inverse become:
$$
g_{\mu \nu} = \left(
\matrix{
-1 & 0 \cr
0 & \ga_{ij}
}
\right), \quad
g^{\mu \nu} = \left(
\matrix{
- 1 & 0 \cr
0 & \ga^{ij}
}
\right)
\fl{ADM4MetricSimp}
$$
and the ADM evolution Einstein equations $\uline{R}_{ji} =0$ become:
$$
A_{\{ij\} \{lm\}} \del_{0} \del_{0} \ga^{lm} + B_{\{ij\}\{mn\}}^{kl} \del_{k} \del_{l} \ga^{mn} \approx 0
\fl{EE}
$$
where we set
$$
A_{\{ij\} \{lm\}} =  \ga_{i(l} \ga_{m) j}
\fn
$$
and:
$$
B_{\{ij\}\{mn\}}^{kl} = -\ga^{kl}\ \ga_{i(m} \ga_{n) j} - \ga_{mn} \de_{i}^{(k}\de_{j}^{l)} 
+ \ga_{j(n} \de^{(k}_{ m)} \de_{i}^{l)} + \ga_{i(m} \de_{ n)}^{(k}\de_{j}^{l)}
\fn
$$
where the symbol $\approx$ means modulo lower order terms with respect to the derivative degree. 
See \ref{ADM} for further details.

Let us stress that the other Einstein equations $\uline{R}_{0j} = 0$ and $\uline{R}_{00} =0$ do not contain second order time derivatives 
(one can eliminate the term $\de_0 K$ by summing with the trace of the equation $\uline{R}_{ji} =0$) and are hence to be interpreted as constraints on the allowed initial conditions. We shall not consider these constraints here (which contain information about the canonical analysis of the system; see \ref{Ash}). 

Next section is devoted to state the Cauchy Problem for the evolution PDE \ShowLabel{EE}.

\ 

\NewSection{The Cauchy problem for a PDE}

%
%

First of all, let us remind that we are not analysing a generic PDE but quasi-linear systems which come from a Lagrangian.
This means that we have a Lagrangian $L = L(x^\mu, y^i, y^i_\mu) dS$ and we obtain the Euler-Lagrangian equation by standard action variation. After the action 
$$
A_D [\si]= \int_D L(y(x), \del y(x)) dS
\fn$$
has been varied (with fixed boundary conditions) we obtain:
$$
\de A_D[\si]= \int_\Si \al_J(x, y) \de y^J dS= 0
\fn$$
and the relative equations are
$$
\al_J(x, y) = 0.
\fn$$

It is clearly that if equations of motion come from a Lagrangian (by using the principle of least action) they will live in the dual space of the fields deformations $\de y^J$. 
In general, by the geometrical framework for variational calculus (see e.g.~\ref{Book}), one can show that Euler-Lagrange equations are described by a (vertical) bundle morphism 
$$
\E: J^{2} \Lor(M)\arr V^\ast(\Lor(M))\otimes A_m(M))
\fn$$
which exactly expresses this remark. 

Let us briefly review the CP for a first order PDE. 
Although GR has second order equations let us  introduce it for first order and then we shall extend the results to the second order system.
Let us also stress that Einstein equation are quasi-linear, namely they are linear in the highest derivative terms.
In this view, a first order PDE analogous to the one for standard GR for us is written as follows:
$$
\al_{IJ}(x, y) \del_0 y^J + \al_{IJ}^i(x, y) \del_i y^J + \ga_I(x, y) = 0.
\fl{FirstOrder}
$$

In the same way, in order to define a CP, we have to define an initial condition, namely:
$$
\eqalign{
&\al_{IJ}(x, y) \del_0 y^J + \al_{IJ}^i(x, y) \del_i y^J + \ga_I(x, y) = 0\cr
&\qquad y^J(0, x^i) = f^J(x^i)
}
\fn$$
where $f^J(x^i)$ is called {\it initial conditions} or {\it initial data}.

We are ready to state the existence and uniqueness theorem:

\THEOREMl{Th}
Let $\al_{IJ}$ be a non-degenerate positive-definite bilinear form and let $\al^i_{IJ}$ be symmetric in indices $IJ$ for all $i$, then
under these hypotheses, given the initial data in $H^k(M)$, with $k > \Frac{n}{2} + 1$, the existence and uniqueness is ensured {in an open interval $I\subset \R$ } and in a suitable Sobolev space, which depends on the regularity of the initial data (see \ref{TAY3}).
\ENDTHEOREM
\ss

We stress that, under the above conditions, if we have $f^J$ smooth, we will have a smooth solution defined on $I$ subset of the real line that contains $t=0$. In other words, we imposed initial conditions and the equations determine a unique solution in a neighbourhood of the Cauchy surface $\Si_0$.

Overviewing technical details, we can notice that two aspects are involved in the theorem above: an analytical and an algebraic condition.
Although they are both important we shall focus on the algebraic one, also in view of the fact that in most contexts
physical fields are chosen to be smooth.

Let us remark that the theorem above states that the well-poseness of the CP is subjected to some algebraic conditions of the coefficients appearing in the differential operator.

Now, we can consider the case of second order quasi-linear systems. 
As done in $\EqRef{FirstOrder}$ we define a second order system as follows:
$$
\matrix{
\al_{IJ}(x, y) \del_{00} y^J - \al_{IJ}^i(x, y) \del_{0i} y^J -  \al_{IJ}^{ij}(x, y) \del_{ij} y^J 
+ \ga_I(x, y, d y) = 0.
}
\fl{secondOrder}
$$
and its associated CP (hereafter we drop the coefficients dependence):
$$
\eqalign{
&\al_{IJ} \del_{00} y^J - \al_{IJ}^i \del_{0i} y^J -  \al_{IJ}^{ij} \del_{ij} y^J \approx 0. \cr
&\qquad y^J(0, x^i) = f^J(x^i) ,
\quad \del_0 y^J(0, x^i) = g^J(x^i).
}
\fl{secondOrderCP}
$$
which will be called hereafter {\it CP2}.

Our goal is to transform a second order PDE in a first order system, by introducing auxiliary fields. 
Inspired by the method used for ODE we can define the following new fields:
$$
\cases{
&v^J = \del_0 y^J \cr
&v^J_j = \del_j y^J \cr
}
\qquad\then
\del_0 v^J_i = \del_i v^J
\fn$$
However, if we wish to consider these equations as part of the original system one should notice that they do not correspond to differential operators with values in the dual space of field variations as it was for the original equation.
Then one should introduce some suitable bilinear forms to write them equivalently in the form:
$$
\cases{
&\be_{IJ} \(\del_0 y^J - v^J \)= 0 \cr
&\be_{IJ}^{ij} \( \del_0 v^J_j - \del_j v^J \) = 0
}
\qquad\then \be_{IJ}^{ij}\(v^J_j - \del_j y^J\)=0 
\fn$$
for some invertible coefficients $\be_{IJ}$ and $\be_{IJ}^{ij}$.  

Let us remark that the equation $v^J_j = \del_j y^J $ contains no time derivative and as such is a constraint on initial conditions and will not contribute to the CP.

Then the equation  \EqRef{secondOrder}  can be written in terms of the new fields $(y^I, v^I, v^I_i)$, so that the CP can be recast in the following form:
$$
\eqalign{
&\cases{
&\be_{IJ} \del_0 y^J \approx 0\cr
&\al_{IJ} \del_{0} v^J - \al_{IJ}^i \del_i v^J - \al_{IJ}^{ij} \del_{i} v_j^J \approx 0\cr
&\be_{IJ}^{ij} \del_0 v^J_j - \be_{IJ}^{ij} \del_j v^J  = 0
}\cr
&\qquad y^J(0, x^i) = f^J(x^i) ,\quad
 v^J(0, x^i) = g^J(x^i), \quad
 v^J_i(0, x^i) = \del_i f^J(x^i)
}
\fl{firstOrderReduced}$$
together with the constraint $\del_i y^J = v_i^J$.
We can write this system in the block-matrix form as follows:
$$
\(
\matrix{
\be_{IJ} &        0       &            0 \cr
    0         &   \al_{IJ} &            0 \cr
    0         &         0       &   \be_{IJ}^{ij} 
}
\)
\del_0 \(
\matrix{
y^J \cr
v^J \cr
v^J_j
}
\)
-
\(
\matrix{
    0         &           0              &           0               \cr
    0         &    \al^k_{IJ}        &   \al_{IJ}^{kj}     \cr
    0         &    \be_{IJ}^{ik}  &    0
}
\)
\del_k \(
\matrix{
y^J \cr
v^J \cr
v^J_j
}
\)
\approx 0
\fl{Eq1}$$

This is a first order CP so the theorem \ShowLabel{Th} applies to it. One has existence and uniqueness of solutions if
the first matrix is symmetric, non-degenerate,  positive-definite  and the second is symmetric.

We already know that $\al_{IJ}$ is non-degenerate and positive-definite. 
If also $\be_{IJ}$ and $\be_{IJ}^{ij} $ are non-degenerate and positive-definite then the whole matrix is.
Since we are free to choose $\be_{IJ}$ as we wish (provided that the choice is non-degenerate and positive-definite)
we can fix it as $\be_{IJ}= \al_{IJ}$, which is automatically a good choice.

For the second matrix to be symmetric $\al^k_{IJ}$ must be symmetric and one must have  (see Appendix $C$)
$$
\be_{IJ}^{ij} =  \al_{JI}^{ij}
\fl{SymmetricMatrix}$$
Thus the block $\be_{IJ}^{ji}$ (and as a consequence of this choice the coefficient $\al_{IJ}^{ij}$)  must be symmetric in $(IJ)$
and non-degenerate positive-definite.

We can rewrite the original system as:
$$
\eqalign{&
\(
\matrix{
\al_{IJ} &        0       &            0 \cr
    0         &   \al_{IJ} &            0 \cr
    0         &         0       &   \al_{IJ}^{ij} 
}
\)
\del_0 \(
\matrix{
y^J \cr
v^J \cr
v^J_j
}
\)
-
\(
\matrix{
    0         &           0              &           0               \cr
    0         &    \al^k_{IJ}        &   \al_{IJ}^{kj}     \cr
    0         &    \al_{IJ}^{ki}  &    0
}
\)
\del_k \(
\matrix{
y^J \cr
v^J \cr
v^J_j
}
\)
\approx 0\cr
&\qquad y^J(0, x^i) = f^J(x^i) ,\quad
 v^J(0, x^i) = g^J(x^i), \quad
 v^J_i(0, x^i) = \ h^J_i(x^i)
}
\fl{RS}$$
which, together with the constraint $\del_i y^J = v_i^J$,  is called the {\it reduced CP} or {\it CP1} for short.

Let us remark that once again, also for second order operators, the well-poseness of the CP is subjected to algebraic requirements.
Unlike for first order operators symmetry is no longer enough and one needs to require that  the coefficient $\al_{IJ}^{kj}$ is also positive-definite.

Now we have to show that the original {\it CP2} (namely \ShowLabel{secondOrderCP}) is dynamically equivalent to the reduced CP \ShowLabel{RS}, namely {\it CP1}. 

Obviously, if we have a solution $y^J(t,x^i)$ of {\it CP2} then:
$$
\(
y^J, v^J := \del_0 y^J, v^J_i := \del_i y^J
\)
\fn
$$
is a solution of the reduced {\it CP1}.
In fact, one immediately has that
$$
\al_{IJ} (v^J- \del_0 y^J)=0
\qquad
\al^{ij}_{IJ}(\del_0 v^J_j-\del_j v^J)=0
\fn$$
while the second order equation can be recast as 
$$
\al_{IJ} \del_0 v^J - \al_{IJ}^i \del_i v^J -\al_{IJ}^{ij}\del_i v^J_j \approx 0
\fn$$
Thus the equations of {\it CP1} are satisfied, the constraint is satisfied (since we defined $v^J_i := \del_i y^J$)
and the constraint, evaluated at $t=0$, shows that $h^J_j(x)= \del_j f^J(x)$ are the only initial conditions compatible with the constraint.
Thus {\it CP1} holds true.

Viceversa, we need to prove that, given a solution $(y^J(t, x), v^J(t, x), v^J_i (t, x))$ of {\it CP1} which satisfies the constraint, 
then $y^J(t, x)$ is also a solution of {\it CP2}.
One has that $v^J =\del_0 y^J$, by first equation in {\it CP1}, and that $\del_0 v^J_j=\del_j v^J= \del_{0j} y^J$, by the third equations.
Thus, one has $ v^J_j= \del_{j} y^J + k(x)$ though, because of the constraint, $k(x)=0$.
Accordingly, one also has 
$$
v^J_j= \del_{j} y^J
\qquad
v^J =\del_0 y^J
\fn$$
with which the second equation of {\it CP1} implies the second order equation.

One obviously has that $y^J(0, x) = f^J(x)$. Now, the first equation of $\EqRef{RS}$ tells us that:
$$
 \del_0 y^J(t, x) =v^J(t,x) 
\quad\then
\del_0 y^J(0, x) = v^J(0,x)= g^J(x)
\fn$$
which is the second initial condition and {\it CP2} holds true.

Obviously the correspondence
$$
(y^J(t, x))
\qquad \longleftrightarrow\quad
\(y^J, v^J := \del_0 y^J, v^J_i := \del_i y^J\)
\fn$$
sends solutions of {\it CP2} into solutions of {\it CP1} and is a bijection, proving dynamical equivalence. 

We have eventually to prove that the if constraint is satisfied at the initial time, then it will be satisfied at all time.
For, let us define the quantity
$$
k^J_j := v^J_j - \del_j y^J
\fn
$$

Since we know that $\del_0 y^J = v^J$, then:
$$
\del_0 k^J_j = \del_0 v^J_j - \del_{0j}  y^J =\del_0 v^J_j - \del_{j} v^J = 0.
\fn$$
where the last equality holds true 
by the last equation of {\it CP1}.

In particular, by imposing the constrain is $t=0$, we obtain $k_j^J(0, x) = 0$ and then we have the following Cauchy problem:
$$
\eqalign{
&\al_{IJ}\del_0 k^J_j = 0\cr
&\qquad
k_j^J(0, x) = 0
}
\fn
$$
which has a solution $k_j^J(t, x^i) = 0$, which is unique since the system is symmetric hyperbolic. 
This means $y^J_j(t,x) = \del_j y^J(t, x)$ at any time and the constraint is satisfied at all times.

Next section is devoted to apply the theory of PDEs developed above to the GR case.

\NewSection{Harmonic coordinates}

We can start to reduce the system \EqRef{EE} of second order to a first order one. We have to define some new fields, as done in the previous section:
$$
\cases{
&\Si^{ij} := \del_0 \ga^{ij} \cr
&\Si^{ij}_l := \del_l \ga^{ij}
}
\fn
$$
so that the system can be rewritten in the following way (as done in the previous section):
$$
\cases{
&A_{\{ij\} \{lm\}} \del_0 \ga^{lm} - A_{\{ij\} \{lm\}} \Si^{lm} = 0 \cr
&A_{\{ij\} \{lm\}} \del_{0}  \Si^{lm} + B_{\{ij\}\{mn\}}^{kl} \del_{k} \Si_l^{mn} \approx 0\cr
&B_{\{ij\}\{mn\}}^{kl} \del_0 \Si^{mn}_k - B_{\{ij\}\{mn\}}^{kl}\del_k \Si^{mn} = 0 
}
\fn
$$
with the constraint
$$
\Si^{ij}_l = \del_l \ga^{ij}.
\fn$$

In matrix form the system above can be written as follows:
$$
\(
\matrix{
A_{\{ij\} \{lm\}}  &        0       &            0 \cr
    0         &   A_{\{ij\} \{lm\}}  &            0 \cr
    0         &         0       &    B_{\{ij\}\{mn\}}^{kl}
}
\)
\del_0 \(
\matrix{
\ga^{lm} \cr
\Si^{lm} \cr
\Si^{mn}_k
}
\)
-
\(
\matrix{
    0         &           0              &           0               \cr
    0         &          0               &     B_{\{ij\}\{mn\}}^{kl}    \cr
    0         &    B_{\{ij\}\{mn\}}^{kl}   &    0
}
\)
\del_k \(
\matrix{
\ga^{lm} \cr
\Si^{mn} \cr
\Si^{mn}_l
}
\)
\approx 0.
\fl{RGReduced}
$$

We want to verify whether the first order system \EqRef{RGReduced} is symmetric hyperbolic. We have two conditions that have to be satisfied, first one is that matrix
$$
{\bf B}=
\(
\matrix{
    0         &           0              &           0               \cr
    0         &          0               &      B_{\{ij\}\{mn\}}^{kl}    \cr
    0         &     B_{\{ij\}\{mn\}}^{kl}   &    0
}
\)
\fn
$$
has to be symmetric which  is automatically satisfied by construction. Second one is that matrix

$$
{\bf A}=
\(
\matrix{
A_{\{ij\} \{lm\}}  &        0       &            0 \cr
    0         &   A_{\{ij\} \{lm\}}  &            0 \cr
    0         &         0       &    B_{\{ij\}\{mn\}}^{kl}
}
\)
\fn
$$
has to be symmetric and positive-definite. One can easily show that the block $A_{\{ij\} \{lm\}}$ is positive-definite  and symmetric.
On the contrary $B^{kl}_{\{ij\} \{lm\}}$ is not  symmetric with respect to the exchange of the pairs $\{ij\}\{lm\}$.

In fact, it is easy to see that its antisymmetric part is:
$$
B^{kl}_{[\{ij\} \{lm\}]} = \Frac{1}{2} \( \ga_{mn} \de^{(k}_i  \de^{l)}_j - \ga_{ij} \de^{(k}_m  \de^{l)}_n \).
\fl{AntyPart}
$$
and it does not  generically vanish. 
Let us also remark that it is a tensor on the space manifold $\Si$ so that one cannot hope it will vanish in any (spatial) coordinate system.

Let us introduce a new coordinates system, called {\it (spatial)  harmonic coordinates}, and defined by the following conditions:
$$
^3\Ga^{l} = \ga^{ij} \GaT{l}{i}{j} = 0
\fl{HCDef}
$$
where $\GaT{l}{i}{j}$ are the Christoffel symbols of $\ga_{ij}$. It is easy to prove that harmonic coordinates always exist.
The condition \EqRef{HCDef} is equivalent to:
$$
\Frac{1}{2} \ga^{kl}\del_{i} \ga_{kl} = \ga^{kl}\del_{l}\ga_{ik}
\fl{HarmoMetric}
$$
which in turn implies
$$
\Frac{1}{2} \ga_{jn} \del_{l} \del_{i} \ga^{jn} \approx \ga_{ij} \del_{l} \del_{n} \ga^{nj}
\fn$$
Then we have
$$
\ga_{jm}\del_{in}\ga^{mn} + \ga_{im}\del_{jn}\ga^{mn}\approx \ga_{mn}\del_{ij} \ga^{mn}
\fn$$
so that, in harmonic coordinates, the coefficient $B_{\{ij\}\{mn\}}^{kl}$ takes the form:
$$
\tilde B_{\{ij\}\{mn\}}^{kl} =\ga^{kl} \ga_{i(m} \ga_{n)j}= \ga^{kl} A_{\{ij\}\{mn\}}
\fn$$
which is  symmetric (as well as automatically non-degenerate positive-definite  since $A_{\{ij\}\{mn\}}$ is. 

In other words, in harmonic coordinates  the antisymmetric part of the full operator becomes lower order, it contributes to lower order tail
and consequently the operator becomes symmetric hyperbolic and CP is well-posed.

Let us remark that there is no contradiction between $B_{[\{ij\}\{mn\}]}^{kl}$ being a tensor and it becoming symmetric in harmonic coordinates
since what we are really saying is that in harmonic coordinates one has
$$
B_{[\{ij\}\{mn\}]}^{kl}\del_{kl}\ga^{mn}\approx 0
\fl{constr}$$
not that the coefficients  $B_{[\{ij\}\{mn\}]}^{kl}$ become zero.
In fact, the equations \ShowLabel{constr} are not covariant and they can be satisfied in particular coordinate systems (e.g.~harmonic coordinates)
without being satisfied in others.

Now that we have seen that harmonic coordinates make evolution equation to be symmetric hyperbolic (though of course they spoil general covariance), we shall show that coordinates more general than harmonic coordinates exist for which the evolution equations are still symmetric hyperbolic.

Since we have already found the antisymmetric part \ShowLabel{AntyPart}, we can directly impose that
$$
B_{\[\{ij\} \{mn\}\]}^{kl} \del_{kl} \ga^{mn}\approx 0
\fl{2har}$$ 
without imposing conditions on first derivatives as in \ShowLabel{HarmoMetric}.

This condition is weaker than harmonic coordinates conditions. 
Obviously, harmonic coordinates imply \ShowLabel{2har}. 
Viceversa, we shall see that there exists a coordinate system in which evolution equations become symmetric hyperbolic and this system is not harmonic;
see \ref{Lusanna}. In fact, the condition \ShowLabel{2har} is also satisfied if one simply has
$$
\del_{k} \del_{l} \ga^{mn}=0
\fl{Vanish}
$$
Indeed, if the inverse metric $\ga^{ij}$ takes a  linear form in coordinates, e.g.
$$
\ga^{-1} =
\(
\matrix{
1 & 0 & 0 \cr
0 & 1 & 0 \cr
0 & 0 & 1+t
}
\)
\fn$$
then \ShowLabel{Vanish} vanishes, while $\Ga^{l} = \ga^{ij} {}^{3}\Ga^{l}{}_{ij}$ does not. 

\NewSection{Bianchi identities and constraints}

In this Section we shall see how one can use projectors (see appendix A) to prove that evolution preserves the constraints
$$
\cases{
&H :=  - \Frac{1}{N}\left( \de_0 K - D_i D^i N \right) - K^{ij} K_{ij} = 0 \cr
&M_i:= D_l \left( K_{j.}{}^l  - \de_j^l K \right)  = 0 \cr
}
\fn$$
due to (contracted)  Bianchi identities.
In fact, Bianchi identities are
$$
\na_\mu G^{\mu \nu} = 0
\fn$$
which by projection provide the following conditions
$$
\cases{
\na_\mu G^{\mu \nu} n_\nu = 0 \cr
\na_\mu G^{\mu \nu} \si^\al_\nu = 0
}
\iff
\cases{
&\de_0 H = - D_i (N M^i) - M^i D_i N - 2 N K H \cr
&\de_0 M^i = - D^i(NH) - 2NM^j K_{j.}{}^i - N K M^i - HD^i N
}
\fl{BianchiIdentities}
$$

By setting, as done above, $N = 1$ and $\vec N = 0$ and we obtain:
$$
\cases{
&\del_0 H \approx - \de^k_i\del_k M^i  \cr
&\ga_{in}\del_0 M^n \approx - \de_i^k\del_k H
}
\iff
\(\matrix{
1 & 0\cr
0 & \ga_{in}
}\)
\del_0\(\matrix{
 H\cr
 M^n }\)
+ \(\matrix{
0 & \de_i^k\cr
\de_i^k & 0 
}\)
\del_k\(\matrix{
 H\cr
 M^i }\)\approx 0
\fn$$
which is symmetric hyperbolic. Then,  a unique solution exists for any initial condition.
If we set $H_{\vert t = 0} = 0$ and $M^i_{\vert t = 0} = 0$ for the initial condition, 
then $H = M^i = 0$ at all $t \in I$ is a solution and since equations are symmetric hyperbolic it is  the only solution. 
Thus the  constraints are preserved by Bianchi identities.

\NewSection{Conclusions and perspectives}

We have seen that (vacuum) Einstein equations can be split into an (elliptic) system of constraints (4 equations in $\dim(M)=4$)
and an evolutionary  system (6 equations in $\dim(M)=4$).

The evolutionary part is not symmetric hyperbolic in general. 
However, one can split the evolutionary part into a symmetric hyperbolic equation and a further constraint, which is non-covariant with respect to change of coordinates on the spatial manifold $\Si$.
Hence one can find spatial coordinates for which the antisymmetric part vanishes and, in that coordinates, solving the symmetric evolutionary part of equations.

The evolutionary equation is covariant with respect to change of coordinates on $\Si$, thus if a unique solution is found in a coordinate system, then a solution is found in any coordinate system.
This is not strange after all since being symmetric hyperbolic is a sufficient (not necessary) condition for solving CP.

Thus there exists a  solution to the evolutionary part for any initial condition. However, not all initial conditions are the same: there are initial conditions which satisfy the elliptic constraints
as well as initial conditions which do not. For any initial condition which satisfies the elliptic constraints one can find a spatial metric $\ga_{ij}(t, x)$, which together with a choice of  the lapse $N$ and shift fields
$\vec N$ (which in fact fixes the ADM foliation),  defines a global Lorentzian metric $g$ which solves original Einstein equations.

This is more or less well known since it is the basis for numerical gravity (see \ref{Gourgoulhon}, \ref{NG}, \ref{Rezzolla}).
However, a detailed analysis allows to clarify some of the details which are relatively less well known and to draw some conclusions which, to the best of our knowledge, are new.

First we clarify the role of (spatial) harmonic coordinates. The coefficients $A_{\{ij\} \{mn\}}$  and $B_{\{ij\} \{mn\}}^{kl}$ of ADM splitting of Einstein equations (i.e.~their principal symbols) are spatial tensors.
Hence, one can prove that changing the coordinates will not make them symmetric if they are not in the first place. 

This originally appeared as an issue to us: how can covariant CP be well posed in a coordinate system and not well posed in another?
The solution is 	quite simple: the evolutionary system is well posed in any coordinate system, just in some coordinate system one can use the theorem about symmetric hyperbolic systems 
while in other coordinates the same system is not symmetric hyperbolic (though the CP is well posed anyway).
In other words, it is the condition of being symmetric hyperbolic that is not covariant.
Nevertheless, the Cauchy theorem for symmetric hyperbolic systems, despite being not covariant, seems to be enough to deal with standard GR in full generality.

As a first issue then, one should define an {\it integrable evolutionary system} as one which is symmetric hyperbolic in at least some coordinate system.

Changing coordinates does not make the system symmetric hyperbolic. It simply makes their antisymmetric part identically satisfied. In other words, the antisymmetric part of the coefficient $B_{[\{ij\} \{mn\}]}^{kl}$
is an remains non-zero. However the equation from it 
$$
B_{[\{ij\} \{mn\}]}^{kl} \del_{kl} \ga^{mn}\approx 0
\fn$$
becomes identically satisfied. This is possible precisely because  this equation is non-covariant.

Moreover, in view of a generalisation, we can stress that the fact that the original Einstein equations come from a variational principle play a fundamental role,
in particular in GR when a metric on fields (namely, $A_{\{ij\} \{mn\}}$)   is unknown since it depends on the unknown field to be determined $\ga_{ij}$.
In particular, one has as many equations as fields. If $k$ equations are constraints then it is reasonable to expect that $k$ fields will be left undetermined, 
eventually spoiling uniqueness as we know it must be in view of the hole argument
(see \ref{Norton}, \ref{PhysState}, \ref{LHole}). 
Moreover, constraints are responsible for the fact that the system is overdetermined.
This simply accounts for the fact that equations of physics are at the same time overdetermined and underdetermined, due to gauge symmetries, see \ref{Gotay1}, \ref{Gotay2}, \ref{Gotay3}, \ref{Gotay4}.
The variational origin of equations also lead us to assume that equations live in the dual space of field deformations, which leads us to a framework in which the metric on fields is not needed 
to define symmetric hyperbolic systems.

Further investigations are needed to generalise this to more general models of interest for  (fundamental) physics. 
It seems possible that integrability is generically equivalent to Hamiltonian formulation, though in a completely Lagrangian setting.

We also found a more general class of coordinate systems than harmonic coordinates for which the system becomes symmetric.
This is not that important in vacuum gravity since harmonic coordinates always exist and they are sufficient to solve the CP.
However, when gravity is coupled to matter fields, matter field equations may need to be symmetrised as well.
Unfortunately, the choice of spatial coordinates in a game which can be played only once since matter equations are coupled to gravity equations.
One would need a coordinate system in which the whole system is symmetric, while the symmetrisation of matter equations depends on the matter-gravity coupling.
Having more coordinate systems in which gravity becomes  symmetric hyperbolic may help when coupled with some matter fields.
We still do not have examples of matter fields which can be solved in this way, though it is clear that harmonic coordinates play no distinguished role within the class of coordinates 
which turn the system in a symmetric hyperbolic system.

\NewAppendix{A}{Projectors}


Let us define the projectors for the ADM decomposition. Due to the immersion of the spacelike hypersurface $\Si$ ($\iota \colon\Si\hookrightarrow M$), we have a normal covector:

$$
u = \Frac{1}{3!} \ep^{ABC} \del_A x^\mu \del_B x^\nu \del_C x^\rho \ep_{\mu \nu \rho \si} dx^\si
\fn
$$
and, due to the metric structure, its normal vector:

$$
\vec{u} = g^{\mu \nu} u_\mu \del_\nu.
\fn
$$

Since $\vec{n}$ is not lightlike, we have its normal unit vector $n$, obtained by normalization.

With $n$ we can define a basis adapted to the foliation: let us take $x = \iota(k) \in M$ we can take a quadruple of vectors: $(n, e_A)$ where $e_A$ is a basis in $T_k \Si$.

This set has the propriety that each $e_A$ is orthogonal to $n$, so it is a basis for $T_{\iota(k)} M$. Also we have that the norm of $n$ is $-1$, i.e.:
$$
g(n,n) = -1.
\fn
$$
Now, we can define some maps which allow us to decompone each geometrical object (like tensors, metrics and so on).
These are defined as follows:
$$
\si^\mu_\nu = \de^\mu_\nu + n^\mu n_\nu.
\fl{Projectors}
$$
and it is easy to see that they are idempotents, so they are projectors. Namely, we have:
$$
\si^\mu_\rho \si^\rho_\nu = \si^\mu_\nu.
\fn
$$

Once projectors are defined we can decompose vectors and covectors (and, in general, tensor) in tangent and normal part.

\THEOREM
For all $v = v^\mu \del_\mu \in T_{\iota(k)} \Si$ there exists a decomposition:

$$
v = v_\parallel + v_\perp
\fn
$$
with $v_\parallel \in T_k \Si$.
This decomposition is unique.
\ENDTHEOREM

\PROOF
$$
v = v^\mu \del_\mu = v^\mu \del_\mu = v^\mu \de_\mu^\nu \del_\nu.
$$
We can obtain (by using \EqRef{Projectors}):

$$
\de^\mu_\nu = \si^\mu_\nu - n^\mu n_\nu
$$
so that:
$$
v = v^\mu (\si_\mu^\nu - n^\nu n_\mu) \del_\nu = v^\mu \si_\mu^\nu\del_\nu - v^\mu n^\nu n_\mu \del_\nu = v^\mu \si_\mu^\nu\del_\nu +(- v^\mu n_\mu) n^\nu \del_\nu.
$$
and we have:
$$
v = v_\parallel + v_\perp 
\fn
$$
wit $v_\parallel = v^\mu \si_\mu^\nu\del_\nu$ and $v_\perp = - v^\mu n_\mu n^\nu \del_\nu$.
It is easy to see that $g(v_\perp, e_A) = g(\vec{n}, v_\parallel) = 0$.
\ENDPROOF

A similar result holds for the covectors.
Let $\al$ a covector on $(\iota(\Si))$, namely:
$$
\al = \al_\mu dx^\mu \in T_{\iota(k)}^*(M).
\fn
$$

Then there exist a decomposition of $\al$ (parallel and perpendicular part) and a covector $\be$ on $\Si$ such that:

$$
\iota^*(\al_\parallel) = \be \in T^*_k (\Si).
\fn
$$

Let now $g = g_{\mu \nu} (x) dx^\mu \otimes dx^\nu$ be a metric on $\iota(\Si)$ and we can write its tangent part:
$$
g_\parallel = g_{\al \be} (x) \si^\al_\mu \si^\be_\nu dx^\mu dx^\nu.
\fn
$$

Now, we can take the pullback along the immersion:
$$
\eqalign{
\iota^*(g_\parallel) = &g_{\al \be} (x(k))\si^\al_\mu \si^\be_\nu \del_A x^\mu \del_B x^\nu dk^A \otimes dk^B =  \cr
g_{\al \be}(\de^\al_\mu + n^\al n_\mu) &(\de^\be_\nu + n^\be n_\nu) \del_A x^\mu \del_B x^\nu dk^A \otimes dk^B = \cr
(g_{\mu \nu} + n_\mu n_\nu) &\del_A x^\mu \del_B x^\nu dk^A \otimes dk^B = g_{\mu \nu} \del_A x^\mu \del_B x^\nu dk^A \otimes dk^B = \cr
\ga_{AB} & dk^A \otimes dk^B = \ga
}
\fn
$$

This means that the parallel part of the metric restricts to the induced metric over $\Si$.

\NewAppendix{B}{}

Let us here consider an example for the situation discussed in Section $2$. 
Fix two fields $y^1$ and $y^2$ and one spatial coordinate on the space $\Si$ (i.e.~$\dim(M)=2$).
In this simple case equation \ShowLabel{Eq1} reads as
$$
\(
\matrix{
\be_{11} & \be_{12} &    0        &       0     & 0 & 0  \cr
\be_{21} & \be_{22} &    0        &       0     & 0 & 0  \cr
    0        &      0       & \al_{11}  & \al_{12} & 0 & 0  \cr
    0        &      0       & \al_{21}  & \al_{22} & 0 & 0  \cr
    0        &      0       &      0       &      0      & \be^{11}_{11} & \be^{11}_{12}    \cr
    0        &      0       &      0       &      0      & \be^{11}_{21} & \be^{11}_{22} 	\cr
}
\)
\del_0
\(
\matrix{
y^1 \cr
y^2 \cr
v^1 \cr
v^2 \cr
v_1^1 \cr
v_1^2 \cr
}
\)
-\(
\matrix{
0 & 0 &    0        &       0     & 0 & 0  \cr
0 & 0 &    0        &       0     & 0 & 0  \cr
    0        &      0       & \al^1_{11}  & \al^1_{12}  & \al^{11}_{11} & \al^{12}_{11} \cr
    0        &      0       & \al^1_{21}  &  \al^1_{22} & \al^{11}_{21} & \al^{12}_{22}  \cr
    0        &      0       &       \be^{1 1}_{11}      &  \be^{11}_{12} & 0   &  0    \cr
    0        &      0       &       \be^{11}_{21}      &  \be^{11}_{22} & 0   &  0   \cr
}
\)
\del_1
\(
\matrix{
y^1 \cr
y^2 \cr
v^1 \cr
v^2 \cr
v_1^1 \cr
v_1^2 \cr
}
\)\approx0
\fn$$
from which one sees that in order to have a symmetric hyperbolic system one needs to verify condition \ShowLabel{SymmetricMatrix}.

Let us stress that the symmetry of the system relies on a suitable ordering of fields and equations. This is acceptable since symmetric hyperbolic form is {\it sufficient},
not a {\it necessary}, condition for having existence and uniqueness of solutions.

\Acknowledgements
We wish to thank Sandro Coriasco, Marco Ferarris, and Mauro Francaviglia for comments and discussion.
We acknowledge the contribution of INFN (Iniziativa Specifica QGSKY), 
the local research project {\it Metodi Geometrici in Fisica Matematica e Applicazioni} (2014) of Dipartimento di Matematica of University of Torino (Italy). 
This paper is also supported by INdAM-GNFM.
 
\ShowBiblio

\end